%% file: main.tex
\documentclass[12pt, draftclsnofoot, onecolumn]{IEEEtran}
\usepackage{balance}

\input{settings}

\input{commands/commands.tex}

\def\endthebibliography{%
  \def\@noitemerr{\@latex@warning{Empty `thebibliography' environment}}%
  \endlist
}
\renewcommand{\baselinestretch}{1.36}

\begin{document}

\IEEEtitleabstractindextext{
\input{abstract}

}
\maketitle
\IEEEdisplaynontitleabstractindextext
\input{sections/introduction.tex}

\input{sections/system_model.tex}

\input{sections/transmission_scheme.tex}

\input{sections/problem_formulation.tex}

\input{sections/alternatingMinimization.tex}
\input{sections/convergence_behaviour.tex}

\input{sections/conclusion.tex}
\input{content/laststuff}
\footnotesize
\bibliographystyle{IEEEtran}
\bibliography{references}
\balance
\input{content/acronyms}

\end{document}

%% file: settings.tex
\usepackage{standard}
\usepackage{etex}

\usepackage{rotate}
\usepackage{algorithmic}
\usepackage{algorithm}
\usepackage{float}
\usepackage{tikz}
\usepackage{refcount}

\usepackage{pgf,tikz}
\usepackage{pgfplots}
\usepackage{pgfplotstable}
\usepackage{bbm}
\usepackage{adjustbox}
\usetikzlibrary{calc}
\usepackage{relsize}
\usepackage{ifthen}

\usetikzlibrary{calc,fit,arrows,plotmarks,intersections,patterns,shapes,decorations,positioning, backgrounds, matrix}

\usepackage[numbers,sort&compress]{natbib}
\usepackage[ngerman, english]{babel}

\usetikzlibrary{arrows.meta}
\edef\wdArrowLength{2}
\tikzset{>={Latex[width=1.5mm,length=\wdArrowLength mm]}}
\usepackage{graphicx}

\usepackage{cancel}
\usepackage{mathtools}
\usepackage{shadethm}
\usepackage{empheq}
\usepackage{skull}
\usepgfplotslibrary{units}
\usepackage{calc}
\usepackage{nicefrac}
\usepackage{fancybox}
\usepackage{psfrag}
\usepackage{dsfont}

\newtheorem{prop}{Proposition}
\title{Synergistic Benefits in IRS- and RS-enabled C-RAN with Energy-Efficient Clustering
\thanks{Part of this work has been published at the IEEE International Conference on Communications (ICC), 2021 \cite{weinberger2020synergistic}.

\noindent K. Weinberger, A. A. Ahmad and A. Sezgin are with the Department of Electrical Engineering Ruhr-University Bochum, Germany (Email: \{kevin.weinberger, alaa.alameerahmad, aydin.sezgin\}@rub.de). A. Zappone is with the University of Cassino and Southern Lazio, Cassino,
Italy (Email: {alessio.zappone}@unicas.it).

This work was funded in part by the Federal Ministry of Education and Research (BMBF) of the Federal Republic of Germany (F\"orderkennzeichen 16KIS1235, MetaSEC).}} 
\author{Kevin Weinberger*, Alaa Alameer Ahmad*, Aydin Sezgin*, Alessio Zappone$^\dagger$}
\date{\today}

\usepackage{graphicx}
\usepackage{placeins}

\usepackage{booktabs}
\usepackage{marvosym}
\usepackage{multirow}

\usepackage{latexsym, amsmath, amssymb, amsfonts, upgreek}

\usepackage[nolist]{acronym}

\usepackage{tikz}
\usetikzlibrary{calc,arrows,positioning,decorations,shadows,shapes,fadings,matrix}
\tikzset{>=latex'}
\tikzset{semithick}

\makeatletter
\providecommand{\IfElsePackageLoaded}[3]{\@ifpackageloaded{#1}{#2}{#3}}
\makeatother

\usepackage{subfigure}
\IfElsePackageLoaded{subfig}
	{\usepackage[subfigure]{tocloft}}{	
	\IfElsePackageLoaded{subfigure}
		{\usepackage[subfigure]{tocloft}}
		{\usepackage{tocloft}}
	}

\makeatletter

\def\tikz@delimiter#1#2#3#4#5#6#7#8{%
	\bgroup
		\pgfextra{\let\tikz@save@last@fig@name=\tikz@last@fig@name}%
		node[outer sep=0pt,inner sep=0pt,draw=none,fill=none,anchor=#1,at=(\tikz@last@fig@name.#2),#3]
		{%
			{\nullfont\pgf@process{\pgfpointdiff{\pgfpointanchor{\tikz@last@fig@name}{#4}}{\pgfpointanchor{\tikz@last@fig@name}{#5}}}}%
			\delimitershortfall\z@
			\resizebox*{!}{#8}{$\left#6\vcenter{\hrule height .5#8 depth .5#8 width0pt}\right#7$}%
		}
		\pgfextra{\global\let\tikz@last@fig@name=\tikz@save@last@fig@name}%
	\egroup%
}

%% file: commands/commands.tex
\tikzset{hexagon/.code={
	\draw (0,2) -- (-4,0) -- (0,-2) -- (4,0) -- (0,2);
}}

\tikzset{phone/.code={
   \node [rectangle,rounded corners=1.5pt,draw,minimum height=0.6cm, minimum width=0.35cm] at (0,0){};
   \node [rectangle,rounded corners=1.5pt,draw,minimum height=0.5cm, minimum width=0.3cm] at (0,0){};
}}

\makeatletter
\def\cantox@vector#1#2#3#4#5#6#7#8{%
  \dimen@.5\p@
  \setbox\z@\vbox{\boxmaxdepth.5\p@
   \hbox{\kern-1.2\p@\kern#1\dimen@$#7{#8}\m@th$}}%
  \ifx\canto@fil\hidewidth  \wd\z@\z@ \else \kern-#6\unitlength \fi
  \ooalign{%
    \canto@fil$\m@th \CancelColor
    \vcenter{\hbox{\dimen@#6\unitlength \kern\dimen@
      \multiply\dimen@#4\divide\dimen@#3 \vrule\@depth\dimen@\@width\z@
      \vector(#3,-#4){#5}%
    }}_{\raise-#2\dimen@\copy\z@\kern-\scriptspace}$%
    \canto@fil \cr
    \hfil \box\@tempboxa \kern\wd\z@ \hfil \cr}}
\def\bcancelto#1#2{\let\canto@vector\cantox@vector\cancelto{#1}{#2}}
\makeatother

\newcommand{\ifthen}[2]{\ifthenelse{#1}{#2}{}}

\newcommand{\myNorm}[1]{\left\lVert#1\right\rVert}

\newcommand{\sqr}[1]{{#1}^2}
\newcommand{\sbrackets}[1]{\left(#1\right)}
\newcommand{\hbrackets}[1]{\left[#1\right]}
\newcommand{\vectw}{\vect{\omega}}
\newcommand{\vectv}{\vect{v}}
\newcommand{\vectvt}{\tilde{\vect{v}}}
\newcommand{\matV}{\mat{V}}

\newcommand{\vecth}{\vect{h}}
\newcommand{\matH}{\mat{H}}
\newcommand{\vecta}{\vect{a}}
\newcommand{\matM}{\mat{M}}

\newcommand{\cbrackets}[1]{\left\{#1\right\}}
\newcommand{\Tr}[1]{\text{Tr}\left(#1\right)}
\renewcommand{\rank}[1]{\text{rank}\left(#1\right)}

\newcommand{\listequationsname}{List of Formulas}
\newlistof{myequations}{equ}{\listequationsname}

\definecolor{myblue1}{rgb}{0,0,255}
\definecolor{myblue2}{rgb}{65,105,225}
\definecolor{myblue3}{rgb}{70,130,180}
\definecolor{myblue4}{rgb}{176,196,222}

\newcommand{\mytilde}{{\raise.17ex\hbox{$\scriptstyle\mathtt{\sim}$}}}
\newcommand{\naive}{}
\def\naive/{na\"{\i}ve}

\newcommand{\executeiffilenewer}[3]{%
\ifnum\pdfstrcmp{\pdffilemoddate{#1}}%
{\pdffilemoddate{#2}}>0%
{\immediate\write18{#3}}\fi%
}

\pgfkeys{/pgf/fpu}
\pgfmathparse{16383+1}

\pgfkeys{/pgf/fpu=false}

%% file: abstract.tex
\begin{abstract}
The potential of intelligent reflecting surfaces (IRSs) is investigated as a promising technique for enhancing the energy efficiency of wireless networks. Specifically, the IRS enables passive beamsteering by employing many low-cost individually controllable reflect elements. The resulting change of the channel state, however, increases both, signal quality and interference at the users. To counteract this negative side effect, we employ rate splitting (RS), which inherently is able to mitigate the impact of interference.
We facilitate practical implementation by considering a Cloud Radio Access Network (C-RAN) at the cost of finite fronthaul-link capacities, which necessitate the allocation of sensible user-centric clusters to ensure energy-efficient transmissions. Dynamic methods for RS and the user clustering are proposed to account for the interdependencies of the individual techniques. Numerical results show that the dynamic RS method establishes synergistic benefits between RS and the IRS. Additionally, the dynamic user clustering and the IRS cooperate synergistically, with a gain of up to 88\% when compared to the static scheme. Interestingly, with an increasing fronthaul capacity, the gain of the dynamic user clustering decreases, while the gain of the dynamic RS method increases. Around the resulting intersection, both methods affect the system concurrently, improving the energy efficiency drastically.

\end{abstract} 

%% file: sections/introduction.tex
\section{Introduction}
With the recent introduction of solutions based on the \ac{IoT} on various application areas, \ac{B5G} mobile wireless networks are expected to support an abundance of devices, while simultaneously providing a 1000 times capacity increase at a similar or lower power consumption than current cellular systems \cite{EE_intr}. To this end, higher frequencies are utilized at the cost of increased channel attenuation, which requires \acp{BS} to be deployed more densely within each cell. Although this ultra dense deployment of small cells decreases the distance between user and \ac{BS}, resulting in a direct channel of higher quality, additional operating costs are introduced to the network. Moreover, this spacial reuse also increases the intra- and inter-cell interference because of the proximity of the \acp{BS} in neighbouring cells. It follows, that the interference also becomes a limiting factor for achieving high efficiency in modern communication networks.

In order to improve the communication quality under these conditions without increasing the necessary power consumption, this work focusses on enhancing the \ac{EE} of the network cost-effectively with the deployment of an \ac{IRS} \citep{anas_irs,IRS_intr2,IRS_intr4,Zhu_Han_IRS,Saad_IRS}.
%
An \ac{IRS} is a metasurface, consisting of many low-cost reflect elements, which are able to induce real-time changes to the reflected signals. As the induced change of each reflect element to the reflected signal is individually adjustable, the \ac{IRS} enables passive beamsteering, which results in an increased efficiency of the communication network \cite{anas_irs,IRS_intr2,IRS_intr4,Zhu_Han_IRS,Saad_IRS}. In a multiuser network, however, each phase shift, induced at the reflect elements, affects the reflected channel path of each user. This results in suboptimal phase shifts for some users and entails additional interfering links within the network. For this reason, the utilization of the passive beamsteering of the \ac{IRS} in a multiuser scenario introduces an increase in interference at the users.

Under these circumstances, the strategy of \ac{TIN} becomes an unattractive solution, as it is known to be a suboptimal strategy, especially in high interference regimes \citep{RS-Base,Kobayashi,CapacityAydin}. Thus, considering the interference caused by spatial reuse and the \ac{IRS}, we deploy \ac{RS} \cite{ourRS,AlaaRSCMD} as an efficient interference mitigation strategy in the context of our paper. In the \ac{RS} strategy, the message of each user is split into a private part that is supposed to be decoded at the intended user only, and a common part, which can be decoded by a subset of users. Leveraging on this concept of \ac{RS}, receivers can adopt \ac{SIC} \cite{SIC}. This enables the mitigation of not only the intra- and inter-cell interference caused by spatial reuse but also inherently mitigates the increased interference caused by the
improved channel gains of the \ac{IRS}. Note that the use of \ac{RS} alone in a classical multiuser network can achieve gains of up to 97\% in terms of the \ac{EE} \cite{Alaa_EE_RSCMD}, if compared to the baseline scheme of \ac{TIN} \cite{aydin-crystal,5074583}. As it will turn out, with the addition of the \ac{IRS}, a synergistic interaction between the \ac{IRS} and the \ac{RS} technique arises. In fact, the concurrent use of both techniques results in a \ac{EE} gain, which is beyond the sum of the \ac{EE} gains, each technique is able to achieve individually.

To enable the practical implementation of the \ac{RS} strategy for a 5G network, this work utilizes a \ac{C-RAN}, in which a \ac{CP} centrally splits and encodes the users' messages and coordinates the transmissions within the network. Moreover, utilizing the \ac{C-RAN} architecture enables the utilization of \ac{CoMP} transmissions based on user-centric clustering. Thus, each user can potentially be served by multiple \acp{BS}, enhancing the user throughput and \ac{EE} \cite{FH_YU,Comp_Yu}.
However, utilizing this data-sharing strategy \cite{FH-datasharing} can also have a negative impact on the \ac{EE} of the network, as each \ac{BS} is connected to the \ac{CP} via fronthaul links of finite capacity. More precisely, in order to realize the \ac{CoMP} transmissions, the \ac{CP} is required to send the same message to multiple \acp{BS}. This effectively introduces redundancy to the fronthaul, which potentially decreases the total achievable rate within the network. Especially in the fronthaul limited regime, this can have a major impact on the \ac{EE}.
It, therefore, becomes equally important to asses the sets of \acp{BS}, which serve each user, while jointly designating the transmission mode of each user (i.e., private, common, or both) \citep{grpSparce,alaa_greenFH}.

In this work, we aim to obtain the full benefit from the cooperation of the \ac{RS} technique and the \ac{IRS} in order to maximize the \ac{EE} of a fronthaul-constrained \ac{C-RAN}. We intend to achieve this by investigating the coupling of the parameters and their interplay within the context of the user-centric clustering. However, capitalizing on this mutual interaction between these techniques poses difficult challenges, due to the dependencies among the parameters of the individual techniques.


\subsection{Contributions}
In order to deal with the interdependencies between the individual techniques,
this paper considers the problem of maximizing the \ac{EE} of the network subject to per-user \ac{QoS} constraints, per-\ac{BS} fronthaul capacity constraints and per-\ac{BS} transmission power constraints. Moreover, the \ac{IRS} is constrained to only induce phase shifts to the reflected signals. The goal of this optimization is to jointly determine the \ac{CMD} sets of each user, the corresponding clusters of \acp{BS} serving each users common and private message and the associated beamformers and rates of each users private and common message, as well as an efficient alignment of the phase shifts at the \ac{IRS}. To deal with the dependencies among the variables, we propose solving the problem in an alternating fashion with the aim of obtaining more tractable subproblems by decoupling the problem. The main contributions can be summarized as follows:

\begin{itemize}
\item \textit{Distributed Optimization:} This paper considers a distributed approach to solve the \ac{EE} problem. Since the emerging subproblems are still dependent on each other, we propose an algorithm that updates specific control-parameters between the optimizations of the individual subproblems in order to influence the solutions to these subproblems. With the chosen sequence, in which the problems are solved, we have the ability to influence the alternating optimization so that we are able to find a parameter set, which is able to exploit the benefits of the combined utilization of the techniques.
\item \textit{Dynamic Common Message Decoding Set Allocation:} We propose a dynamic procedure that determines efficient \ac{CMD} sets, which cooperate well with the \ac{IRS} as they are updated dynamically throughout the alternating optimization algorithm. Because the effectiveness of the \ac{RS} technique to mitigate interference is based on the \ac{CMD} sets, it is vital that the sets are updated with regards to the phase shifters of the \ac{IRS}.
\item \textit{Dynamic User-Centric Clustering:} The \ac{EE} of the network is dependent on sensible clusters, the choice of which highly depends on the parameters of both, \ac{RS} and the \ac{IRS}. We therefore adopt a formulation of the \ac{EE} problem that enables the dynamic allocation of clusters (for both, the private and common transmissions), which represent the perfect tradeoff between sending redundant messages to multiple \acp{BS} and the consumed network power. To this end, we utilize an $\ell_0$ relaxation technique in combination with an inner-convex approximation framework \cite{AlaaRSCMD} and formulate an objective function that encourages discarding inefficient BS-user links. In fact, the non-convex fronthaul constraint is approximated with a surrogate convex function, which are accounted for by updating the respective values with in an outer-loop iteratively.
\item \textit{Phase Shift Optimization:} In order to guarantee the convergence of the alternating optimization approach, we determine phase shifters, which exclusively improves both, the achievable private rate and common rates at each user. By proposing this approach, we not only ensure the convergence of the alternating optimization, but also enhance the \ac{EE} of the network by improving the achievable rates within the network. 
\item \textit{Numerical Simulations:} Through extensive numerical simulations we are able to gain insights into the interactions between the different techniques. We are able to show the effectiveness of the exploitation of the cooperative benefits by providing an in-depth numerical investigation, in which the gains of the combined techniques are compared to the gains of the individuals schemes as well as the baseline schemes.
\end{itemize}

%% file: sections/system_model.tex
\section{System Model}\label{ch:Sysmod}

\begin{figure}
\centering
\includegraphics[width=0.75\columnwidth]{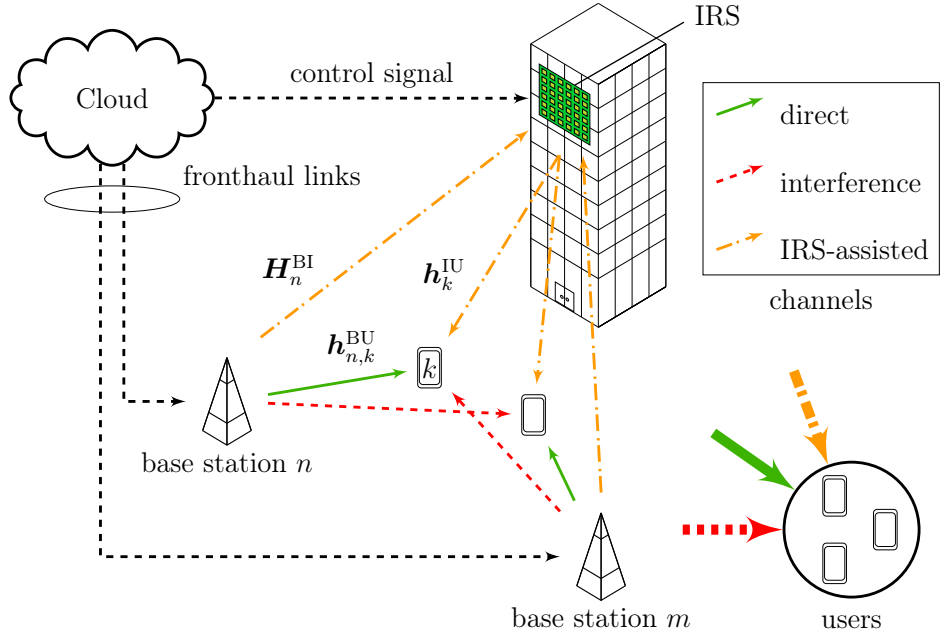}
\caption{IRS-assisted multiuser \ac{C-RAN} system}
\label{fig:IRS-CRAN}
\end{figure}

In this work we consider the system model depicted in Figure \ref{fig:IRS-CRAN}, which consists of an \ac{IRS}-assisted and \ac{RS}-enabled \ac{C-RAN} downlink system. More precisely, the network consists of a set of multi-antenna \acp{BS} $\mathcal{N} = \{1,2,\cdots,N\}$, each of which is equipped with $L \geq 1$ antennas. A set of single-antenna users $\mathcal{K} = \{1,2,\cdots,K\}$ is served by the \acp{BS}. We consider the deployment of an \ac{IRS}, composed of $R$ passive real-time-controllable reflect elements, in the communication environment to support the \acp{BS} in their transmissions to the users. The \ac{CP} of the \ac{C-RAN} is connected to each \ac{BS} $n \in \mathcal{N}$ via orthogonal fronthaul links of finite capacity $C_n$, while the \ac{IRS} is connected to the \ac{CP} via a seperate fronthaul link. Each user $k$ has a \ac{QoS} target, which is represented by a minimum data rate $r_k^{\text{Min}}$.


The direct channel link between \ac{BS} $n$ and user $k$ is denoted by ${\vect{h}_{n,k} \in \mathbb{C}^{L\times1}}$. Moreover, $\mat{H}_{n}^\text{BI} \in \mathbb{C}^{L \times R}$ denotes the channel link from \ac{BS} $n$ to the \ac{IRS} , while $\vecth_{k}^\text{IU} \in \mathbb{C}^{R \times 1}$ denotes the channel link from \ac{IRS} to the user $k$. 
Similarly, we denote $\vect{h}_k = [ ({\vect{h}_{1,k}^{\text{BU}}})^{\,T } ,({\vect{h}_{2,k}^{\text{BU}}})^T, \dots ,({\vect{h}_{N,k}})^T ]^T  \in \mathbb{C}^{NL\times1}$ as the aggregate direct channel vector of user $k$, $\vect{H}^{\text{BI}} = \left[ ({\vect{H}_{1}^{\text{BI}}})^{\,T } ,({\vect{H}_{2}^{\text{BI}}})^T, \dots ,({\vect{H}_{N}^{\text{BI}}})^T \right]^T  \in \mathbb{C}^{NL\times R}$ as the aggregate channel matrix from the \acp{BS} to the \ac{IRS} and $\vect{x} = \left[ {\vect{x}_{1}}^{\,T } ,{\vect{x}_{2}}^T, \dots , {\vect{x}_{N}}^T \right]^T  \in \mathbb{C}^{NL\times1}$ as the aggregate transmit signal vector. Using this notation, the received signal at user $k$ can be written as the sum of the direct and \ac{IRS}-assisted link, namely
\begin{align}\label{eq:recSig}
    y_k & = \vect{h}_{k}^H\, \vect{x} + (\mat{H}^\text{BI} \, \mat{\Theta} \, \vect{h}_{k}^{\text{IU}})^H \, \vect{x} + n_k \, ,
\end{align}
where $n_k \sim \mathcal{CN}(0,\sigma^2)$ is the \ac{AWGN} and $\mat{\Theta}\, = \text{diag}(\vectv) \in \mathbb{C}^{R \times R}$ is a diagonal reflection coefficient matrix representing the response of the reflect elements. We define the phase-shift vector as $\vectv = [e^{j\theta_1},\dots, e^{j\theta_R}]^T$, where each phase shift $\theta_r \in [0,2\pi]$ induced at the $r$-th reflect element is represented by the corresponding reflection coefficient defined as $v_r = e^{j\theta_r}$.
This notation facilitates denoting the combination of the aggregated direct and reflected channel vectors of user $k$ as an effective channel, namely
$\vecth_k^{\text{eff}}(\vectv) = \vect{h}_{k}^{\text{BU}} + \mat{H}_k \vectv$, where $\mat{H}_k = \mat{H}^{\text{BI}} \text{diag}(\vect{h}^{\text{IU}}_k)$. The received signal (\ref{eq:recSig}) at user $k$ can thus be rewritten as $y_k = (\vecth_k^{\text{eff}}(\vectv))^H \vect{x}+ n_k$.

%% file: sections/transmission_scheme.tex
\subsection{Rate Splitting (RS)}
The \ac{RS} scheme can be adopted with different approaches  \citep{RateSplitBruno,ourRS}. The scheme adopted in this paper is based on \cite{ourRS}, in which the \ac{CP} splits the requested message of user $k$, denoted as $q_k$, into two sub-messages, namely a private part $q_k^p$ and a common part $q_k^c$. The \ac{CP} subsequently encodes the respective parts into the private and common symbols denoted as $s_k^p$ and $s_k^c$, respectively. It is assumed that these messages form an \ac{i.i.d.} Gaussian codebook. Afterwards, the \ac{CP} distributes the private symbols $s_k^p$ and common symbols $s_k^c$ with a predetermined cluster of \acp{BS}, that exclusively transmits the beamformed private or common symbols to user $k$. We denote the subsets of users that are served by \ac{BS} $n$ with a private or common message as $\mathcal{K}_n^p,\,\mathcal{K}_n^c \subseteq \mathcal{K}$, respectively.
With the knowledge of these subsets, the \ac{CP} is able to create the beamformers $\vectw_{n,k}^p$ and $\vectw_{n,k}^c$, used by \ac{BS} $n$ to send $s_k^p$ and $s_k^c$, respectively, and forwards them to \ac{BS} $n$ through the fronthaul link $C_n$ along with the respective private symbols $\cbrackets{s_k^p \,|\, \forall k \in \mathcal{K}_n^p}$ and common symbols $\cbrackets{s_k^c \,|\, \forall k \in \mathcal{K}_n^c}$. Due to the finite fronthaul capacity limit $C_n$, the achievable transmission rate is subjected to the following fronthaul constraint:
\begin{align}\label{eq:fh_const_simple}
  \sum_{k \in \mathcal{K}_n^p}R_k^p + \sum_{k \in \mathcal{K}_n^c} R_k^c \leq C_n \, , \forall n \in \mathcal N,
\end{align}
where $R_k^p$ and $R_k^c$ are the private and common rate of user $k$, respectively, and thus $R_k = R_k^p+R_k^c$, where $R_k$  is the rate of user $k$.

After receiving the symbols with the corresponding beamformers, each \ac{BS} $n$ constructs the overall transmit signal vector $\vect{x}_n$, which is  defined as
\begin{align}\label{eq:xvec}
   \vect{x}_n &= \sum_{k\in\mathcal{K}_n^p} \vectw_{n,k}^p s_k^p + \sum_{k\in\mathcal{K}_n^c} \vectw_{n,k}^c s_k^c,
\end{align}
and subject to the following power constraint:
\begin{align}\label{eq:powCons}
    \mathbb{E}\left\{\vect{x}_n^H \vect{x}_n\right\} \leq P_n^{\text{Max}} , \quad \forall n \in \mathcal{N} \, ,
\end{align}
where $P_n^{\text{Max}}$ represents the maximum transmit power that is available at \ac{BS} $n$.
By denoting the aggregate beamforming vectors as $\vectw_k^o = \left[ (\vectw_{1,k}^o)^T,(\vectw_{2,k}^o)^T,\dots,(\vectw_{N,k}^o)^T  \right]^T  \in \mathbb{C}^{NL\times1}, \forall o \in \{p,c\}$ associated with $s_k^o$, the aggregate transmit signal vector can be constructed as \begin{align}\label{eq:transSig}
\vect{x} = \sum_{k \in \mathcal{K}_n^p}\vectw_k^p s_k^p + \sum_{k \in \mathcal{K}_n^c}\vectw_k^c s_k^c.
\end{align}
Using the definition of the transmit signal vector (\ref{eq:transSig}), the power constraint (\ref{eq:powCons}) can be rewritten as
\begin{align}\label{eq:powConsII}
    \sum_{k\in\mathcal{K}} ||\vectw_{n,k}^p ||_2^2 + || \vectw_{n,k}^c ||_2^2 \leq P_n^{\text{Max}} , \quad \forall n \in \mathcal{N} .
\end{align}
Note that if \ac{BS} $n$ does not participate in the cooperative transmission of the private or common symbol of user $k$, the respective beamformers are set to zero, i.e., $\vectw_{n,k}^p = \zero_L$ or $\vectw_{n,k}^c = \zero_L$, where $\zero_L$ denotes a column vector of length $L$ with all zero entries. This can be equivalently expressed in terms of the indicator function \cite{FH_YU}, i.e.,
\begin{align}
\ones \left\{ \norm{\vectw^o_{n,k}}_2^2 \right\} =
    \begin{cases}
        1 \quad ||\vectw^o_{n,k}||_2^2 > 0  \\
        0 \quad \text{otherwise}
    \end{cases},  \forall o \in \{p,c\}.
\end{align}
Without loss of generality, this indicator function can be written as the $\ell_0$-norm, i.e., $\ones \left\{ ||{\vectw^o_{n,k}}||_2^2 \right\} = \norm{||\vectw^o_{n,k}||_2^2}_0$ because the power that BS $n$ transmits to user $k$ is a positive scalar, i.e., $||{\vectw^o_{n,k}}||_2^2 \in \mathbb{R}_+$. Using this definition, the subset of users $\mathcal{K}_n^p$ and $\mathcal{K}_n^c$ can be written as
\begin{align}
   \mathcal{K}_n^p &= \{k\in\mathcal{K} \, | \, \quad \norm{||\vectw^p_{n,k}||_2^2}_0 = 1   \}, \label{eq:subs1_rewritten}\\
   \mathcal{K}_n^c &= \{k\in\mathcal{K} \, | \, \quad \norm{||\vectw^c_{n,k}||_2^2}_0 = 1   \} \label{eq:subs2_rewritten}.
\end{align}
Using the expressions (\ref{eq:subs1_rewritten}) and (\ref{eq:subs2_rewritten}), the fronthaul capacity constraints in (\ref{eq:fh_const_simple}) can be reexpressed in the following form:
\begin{align}\label{eq:CapIndex}
  \sum_{k\in\mathcal{K}} \left(\norm{||\vectw^p_{n,k}||_2^2}_0 R_k^p + \norm{||\vectw^{c_{\phantom{y}}}_{n,k}||_2^2}_0 R_k^c\right) \leq C_n^{\mathsf{max}} \, , \forall n \in \mathcal{N}.
  \end{align}

\subsection{Achievable rates}
%
In this work, the influence of the \ac{CMD} scheme, adopted by the users, is utilized for the purpose of interference mitigation, especially focussing on mitigating the additional interference caused by the \ac{IRS}-enhanced channels.
Each user decodes a subset of common messages with a fixed decoding order in addition to their private messages. It follows, that choosing the decoding order for user $k$, decoding its common and private messages, is vital for the performance of using \ac{RS} as interference mitigation technique.
%
Hence, we adopt a successive decoding strategy, in which user $k$ decodes the common message of the user, whose interference is the strongest, first. This  enables each user to decode and efficiently cancel parts of the received interference.
%
Thus, let $\mathcal{M}_k$ be the set of users decoding $s_k^c$:
\begin{align}
\mathcal{M}_k = \cbrackets{j \in \mathcal{K} \,|\, \text{ user } j \text{ decodes }s_k^c}.
\end{align}
In addition, let the set of users $\Phi_k$, whose common messages are decoded by user $k$, and the set of users $\bar{\Phi}_k$, whose common messages are \textit{not} decoded by user $k$, be defined as
\begin{align}
\Phi_k = \cbrackets{j \in \mathcal{K} \, | \, k \in\mathcal{M}_j},\, \bar{\Phi}_k = \cbrackets{j \in \mathcal{K} \, | \, k \notin\mathcal{M}_j},
\end{align}
in which $\Phi_k$ and $\bar{\Phi}_k$ are two disjoint subsets from the set of active users $\mathcal{K}$, while the cardinality of $\Phi_k$ is bounded by the number of layers in the successive decoding strategy $D$, i.e., $|\Phi_k| \leq D$.
For these sets $\Phi_k ,\, \forall k\in\mathcal{K}$ a decoding order is established, which is represented by a mapping of an ordered set with cardinality of $|\Phi_k|$, i.e.:
\begin{align}
   \pi_k(j) : \cbrackets{1,2,\dots,|\Phi_k|} \rightarrow \Phi_k.
\end{align}
Accordingly, the expression $\pi_k(j_1) > \pi_k(j_2)$ signifies that user $k$ prioritizes decoding the common message of user $j_1$ before decoding the common message of user $j_2$, assuming $j_1\neq j_2$.

Using the expressions above, the received signal at user $k$ can be expressed as
\begin{align}\label{eq:recSig_final}
  &y_k = \overbrace{\sbrackets{\vecth_k^{\text{eff}}(\vectv)}^H \vectw_k^p s_k^p + \sum_{j \in \Phi_k} \sbrackets{\vecth_k^{\text{eff}}(\vectv)}^H \vectw_j^c s_j^c}^{\text{signals that are decoded}}+ \nonumber\\ &   \underbrace{\sum_{m \in \mathcal{K}\text{\textbackslash}\cbrackets{k}} \hspace{-0.35cm} \sbrackets{\vecth_k^{\text{eff}}(\vectv)}^H \vectw_m^p s_m^p \hspace{-0.05cm} +\hspace{-0.05cm} \sum_{\ell \in \bar{\Phi}_k} \sbrackets{\vecth_k^{\text{eff}}(\vectv)}^H \vectw_\ell^c s_\ell^c + n_k}_{\text{interference plus noise}}.
\end{align}
Let $\gamma_k^p$ denote the \ac{SINR} of user $k$ decoding its private message, and let $\gamma_{i,k}^c$ denote the \ac{SINR} of user $i$ decoding the common message of user $k$. Let the stacked private and common beamformers of user $k$  be denoted as $\vectw_k = [ \sbrackets{\vectw_k^p}^T , \sbrackets{\vectw_k^c}^T ]^T \in \mathbb{C}^{2NL\times1}$  and let $\vectw = [ (\vectw_1)^T , (\vectw_2)^T , \dots , (\vectw_K)^T]^T \in \mathbb{C}^{2KNL\times1}$ represent all beamformers. Using the partition done in (\ref{eq:recSig_final}), $\gamma_k^p(\vectw,\vectv)$ and $\gamma_{i,k}^c(\vectw,\vectv)$ can be formulated as
\begin{align}
   \gamma_k^p(\vectw,\vectv) &= \label{eq:gammap}\frac{|{\sbrackets{\vecth_k^{\text{eff}}(\vectv)}^H \vectw_k^p}|^2}
                     {\sum\limits_{m \in \mathcal{K}\text{\textbackslash}\cbrackets{k}} \hspace{-0.2cm} |{\sbrackets{\vecth_k^{\text{eff}}(\vectv)}^H \vectw_m^p}|^2 + \sum\limits_{\ell \in \bar{\Phi}_k} |{\sbrackets{\vecth_k^{\text{eff}}(\vectv)}^H \vectw_\ell^c}|^2 + \sigma^2}, \\
   \gamma_{i,k}^c(\vectw,\vectv) &= \label{eq:gammac}\frac{|{\sbrackets{\vecth_i^{\text{eff}}(\vectv)}^H \vectw_k^c}|^2}
                     {T_i + \sum\limits_{\ell \in \bar{\Phi}_i} |{\sbrackets{\vecth_i^{\text{eff}}(\vectv)}^H \vectw_\ell^c}|^2 +
                     \sum\limits_{m \in \Omega_{i,k}} |{\sbrackets{\vecth_i^{\text{eff}}(\vectv)}^H \vectw_m^c}|^2},
\end{align}
in which $T_i = \sum_{j \in \mathcal{K}} |\sbrackets{\vecth_i^{\text{eff}}(\vectv)}^H \vect{\omega}_j^p|^2  + \sigma^2$ and where the set $\Omega_{i,k}$ represents the set of users whose common messages are decoded by user $i$ after decoding the common message of user $k$, defined as
\begin{align}
   \Omega_{i,k} = \cbrackets{m\in\Phi_i \, | \, \pi_i(k) > \pi_i(m)}.
\end{align}
To assure, that each user is served with its request, the \ac{QoS} $r_k^{\mathsf{Min}}$ of each user $k$ can be defined as
\begin{align}\label{eq:QoS}
   R_k^p + R_k^c \geq r_k^{\mathsf{Min}}, \forall k \in \mathcal{K}\,,
\end{align}
where the messages $s_p^k$ and $s_c^k$ are only decoded reliably if the following conditions are satisfied by the private and common rates of each user
\begin{align}\label{eq:privateRate}
   B \log_2 (1+\gamma_{k}^p (\vectw,\vectv)) \geq& R_k^p , \quad \forall k \in \mathcal{K}\\
   B \log_2 (1+\gamma_{i,k}^c(\vectw,\vectv)) \geq& R_k^c, \quad  \forall k \in \mathcal{K} ,\, \forall i \in \mathcal{M}_k,\label{eq:commonRate}
\end{align}
in which $B$ denotes the transmission bandwidth.


%% file: sections/problem_formulation.tex
\section{Problem Formulation}\label{ch:propform}
\input{sections/problem_formulation/general_problem.tex}

%% file: sections/problem_formulation/general_problem.tex

In this work, we are interested in the joint optimization of the power control,  private and common rate allocations, \ac{CMD} sets, clusters and their associated beamformer design for each user. In addition, we aim to jointly determine the phase shift design for the \ac{IRS}, which maximizes the \ac{EE} of the network subject to per-user \ac{QoS} constraints, shared per-\ac{BS} fronthaul and power constraints.
%
To this end, let the total transmit power $P^{\mathsf{Tr}}$ be defined as:
\begin{align}\label{eq:transmit_tot}
   P^{\mathsf{Tr}}(\vectw) &= \sum_{k\in\mathcal{K}} \sum_{n\in\mathcal{N}}
   ( ||\vectw_{n,k}^p ||_2^2 + || \vectw_{n,k}^c ||_2^2 ).
\end{align}
Let the stacked private and common rate of each user $k$ be denoted as $\mat{R}_k= \hbrackets{R_k^p,R_k^c}^T$ and let $\mat{R}= [\mat{R}_1^T,\dots,\mat{R}_K^T]^T \in \mathbb{R}_+^K$ represent all rates. Let the total rate of the network be defined as:
\begin{align}\label{eq:rate_tot}
   R^{\mathsf{t}}(\mat{R}) &= \sum_{k\in\mathcal{K}}
   ( R_k^p + R_k^c ).
\end{align}
The problem can then be mathematically formulated as
\begin{align}\label{eq:genProb}
   &\underset{\vectw,\vectv, \mat{R},\mathcal{S}}{\text{maximize}} \quad  \frac{R^{\mathsf{t}}(\mat{R})}{P^{\mathsf{Tr}}(\vectw)+P^{\mathsf{Circ}}(\vectw,\mat{R})} = E^{\mathsf{Obj}}(\vectw,\vect{R}) \tag{P1}\\
   &\text{subject to} \quad  (\ref{eq:powConsII}), (\ref{eq:CapIndex}), (\ref{eq:QoS}), (\ref{eq:privateRate}) , (\ref{eq:commonRate}),  \nonumber \\
   & \qquad \qquad \quad |v_r| = 1,\qquad  \forall r \in \cbrackets{1,...,R},\label{eq:general_IRS}
\end{align}
where the unit-modulus constraints in (\ref{eq:general_IRS}) represent the phase shift constraint $0\leq \theta_r \leq 2\pi, \forall r\in \{1,\dots,R\}$ and the variable
\begin{align}
\mathcal{S} = \{ \mathcal{M}_k , \Phi_{k} , \bar{\Phi}_k\}_{k\in\mathcal{K}}
\end{align}
is introduced, which represents the set variables of the \ac{RS} and the successive decoding strategy. Furthermore, $P^{\mathsf{Circ}}(\vectw,\mat{R}) = P^{\mathsf{Pr}} + P^{\mathsf{IRS}} + P^{\mathsf{FH}}(\vectw,\mat{R})$ represents the required processing power of the network. More precisely, it is defined by the signal processing circuitry of the \acp{BS} and the processing power of the \ac{CP}, denoted as $P^{\mathsf{Pr}}$, as well as the required power to perform phase shifting on the impinging signal $P^{\mathsf{IRS}} = NP^\mathsf{IRS}_r$, where $P^\mathsf{IRS}_r$ represents the power consumption of each phase shifter \cite{IRS_intr2}. Moreover, $P^{\mathsf{FH}}(\vectw,\mat{R})$ represents the power that is required for the fronthaul traffic and can be defined as
\begin{align}\label{eq:fronthaul_clust}
P^{\mathsf{FH}}(\vectw,\mat{R}) &= P^{\mathsf{Mbps}}\sum_{k\in\mathcal{K}}\sum_{n\in\mathcal{N}}  \left(\norm{||\vectw^p_{n,k}||_2^2}_0 R_k^p + \norm{||\vectw^{c_{\phantom{y}}}_{n,k}||_2^2}_0 R_k^c\right)
\end{align}
where $P^{\mathsf{Mbps}}$ is an \ac{EE} factor representing the power consumption per bitrate \cite{EE_factor}. The problem is generally difficult to solve, due to the non-convex constraints in combination with the set design problem. Moreover, the optimization variables $\vectw$ and $\vectv$ are coupled in (\ref{eq:privateRate}) and (\ref{eq:commonRate}).

%% file: sections/alternatingMinimization.tex
\input{sections/alternatingMinimization/altMin.tex}
\input{sections/alternatingMinimization/w_problem.tex}
\input{sections/alternatingMinimization/v_problem.tex}
\input{sections/alternatingMinimization/cmd_problem.tex}
\input{sections/alternatingMinimization/obtaining_optimal_solution.tex} 

%% file: sections/alternatingMinimization/w_problem.tex
\section{Alternating Optimization}
To facilitate practical implementation, the problem is solved in an alternating fashion by proposing an alternating optimization framework. This decouples the variables $\vectv$ and $\vectw$ and $\mathcal{S}$ and results in three subproblems originating from (\ref{eq:genProb}), namely a beamforming design problem, a phase shift design problem and a set design problem, which are solved individually with different frameworks. In order to obtain the most efficient combination of \ac{IRS} phase shifters $\vectv$ and set variables $\mathcal{S}$, we propose to optimize the phase shifters to prioritize the sum-path gain in the beginning of the optimization. This results in potentially higher efficiency of the phase shifters at the cost of high interference at some users, which we then mitigate by designing sensible set variables $\mathcal{S}$. Moreover, as elaborated later, the non-smooth $\ell_0$-norm is approximated with a smooth approximation \cite{AlaaRSCMD}. However, even with this approximation, the optimization problem is still defined by a non-convex feasible set. To address this problem, we derive surrogate upper-bound functions for all non-convex functions, thereby approximating the non-convex set with a convex one. The approximations are then iteratively improved until convergence by using the \ac{SCA} approach \cite{Alaa_EE_RSCMD}.
\subsection{Relaxation of the $\ell_0$-norm}
To approximate the non-smooth, non-convex $\ell_0$ norm, we utilize a smooth and convex approximation function. To this end, we consider the function
\begin{align}\label{eq:atan}
    f_\Alpha(x) = \frac{2}{\pi} \text{arctan}\left(\frac{x}{\Alpha}\right) , \quad x\geq0 ,
\end{align}
which is frequently used to approximate the $\ell_0$-norm and where $\Alpha>0$ is the smoothness parameter controlling the quality of the approximation \cite{arctan_approx}. With the help of this function we are able to derive the upper-bound functions for (\ref{eq:CapIndex}) and (\ref{eq:fronthaul_clust}) by replacing the non-convex $\ell_0$-norm with the approximated convex function, namely
\begin{align}\label{eq:ApproxCap}
\sum_{k\in \mathcal{K}} f_\Alpha\left(||\vectw^p_{n,k}||_2^2\right) R_k^p +  f_\Alpha\left(||\vectw^c_{n,k}||_2^2\right) R_k^c \leq C_n \, , \forall n \in \mathcal{N}\,,
\end{align}
\begin{align}\label{eq:fh_fa}
P_\Alpha^{\mathsf{FH}}(\vectw,\mat{R}) = \,\,  P^{\mathsf{Mbps}}\sum_{k\in\mathcal{K}}\sum_{n\in\mathcal{N}}  \left(f_\Alpha(||\vectw^p_{n,k}||_2^2) R_k^p + f_\Alpha(||\vectw^{c_{\phantom{y}}}_{n,k}||_2^2) R_k^c\right).
\end{align}

\subsection{Beamforming Design}
Due to the alternating optimization approach, $\vectv$ and $\mathcal{S}$ are assumed to be fixed during the beamforming design procedure. This facilitates rewriting the optimization problem (\ref{eq:genProb}) as
\begin{align}\label{eq:genW}
    & \hspace{-4cm}\underset{\{\vect{w}_k,\vect{t}_k,\mat{R}_k\}_{k=1}^K}{\text{maximize}} \quad \frac{R^{\mathsf{t}}(\mat{R})}{P^{\mathsf{Tr}}(\vectw)+P_\Alpha^{\mathsf{Circ}}(\vectw,\mat{R})} \tag{P2.1}\\
    & \hspace{-4cm}\text{subject to} \quad  (\ref{eq:powConsII}), (\ref{eq:QoS}) , (\ref{eq:ApproxCap}) \nonumber 
    \end{align}
    \vspace{-0.95cm}
    \begin{align}
    & R_k^p - B \log_2(1+t_k^p) \leq 0 && \forall k \in \mathcal{K},\label{eq:p2.1_first}\\
    & R_k^c - B \log_2(1+t_k^c) \leq 0 && \forall k \in \mathcal{K},\\
   &\vect{t}_k \geq 0&& \forall k \in \mathcal{K},\\
   &\mat{R}_k \geq 0 && \forall k \in \mathcal{K},\label{eq:2.1_last}\\
   & t_k^p \leq  \gamma_k^p(\vectw),     && \forall k \in \mathcal{K} \label{eq:tkp}\\
   & t_k^c \leq  \gamma_{i,k}^c(\vectw), && \forall k \in \mathcal{K} ,\, \forall i \in \mathcal{M}_k, \label{eq:tkc}
\end{align}
where the variables $\vect{t}_k=\hbrackets{t_k^p,t_k^c}^T$ are introduced and $P_\Alpha^{\mathsf{Circ}}(\vectw,\mat{R})$ denotes the utilization of (\ref{eq:fh_fa}) instead of (\ref{eq:fronthaul_clust}). The notation $\vect{t}_k  \geq  0$ and $\mat{R}_k  \geq  0$ indicates that vector $\vect{t}_k$ and matrix $\mat{R}_k$ are greater than or equal to 0 in a component-wise manner.
Problem (\ref{eq:genW}) is a fractional programming problem and is conventionally solved with the Dinkelbach algorithm \cite{dinkelbach}. However, the feasible set of problem (\ref{eq:genW}) is non-convex as the constraints (\ref{eq:tkp}) and (\ref{eq:tkc}) still define a non-convex feasible set, which makes it computationally inefficient to apply the Dinkelbach transformation directly to solve the problem \cite{dinkelbach_ineff}. We overcome this challege by applying the \ac{SCA} approach in combination with the Dinkelbach transformation \cite{Alaa_EE_RSCMD}. In order to obtain a convex representation of the constraints (\ref{eq:tkp}) and (\ref{eq:tkc}), they can be rewritten into the following form%
\begin{align}\label{eq:tkp2}
         & {\sum_{m\in\mathcal{K}\text{\textbackslash}\cbrackets{k}}^{K} |\sbrackets{\vecth_k^{\text{eff}}}^H \vect{\omega}_m^p|^2 + \sum_{\ell \in \bar{\Phi}_k} |\sbrackets{\vecth_k^{\text{eff}}}^H \vect{\omega}_\ell^c|^2 + \sigma^2}\, -\, {\frac{|\sbrackets{\vecth_k^{\text{eff}}}^H \vect{\omega}_k^p|^2}{t_k^p}}\leq 0,\\
         &{T_i + \sum_{\ell \in \bar{\Phi}_i} |\sbrackets{\vecth_i^{\text{eff}}}^H \vect{\omega}_\ell^c|^2 \phantom{\frac{|}{|}} \hspace{-0.075cm} + \sum_{m \in \Omega_{i,k}} |\sbrackets{\vecth_i^{\text{eff}}}^H \vect{\omega}_m^c|^2}\,\  -\, {\frac{|\sbrackets{\vecth_i^{\text{eff}}}^H \vect{\omega}_k^c|^2}
         {t_k^c}} \leq 0 \label{eq:tkc2}    ,
   \end{align}
with an abuse of notation by skipping the dependency of $\vect{h}_k^\text{eff}$ from $\vectv$.
Next, the constraints can be approximated by using the first-order Taylor approximation around a feasible point $(\tilde{\vectw},\tilde{\vect{t}})$ \cite{convexFunct2}, as they are represented by a difference of convex functions.
By applying the first-order Taylor approximations to the fractional terms in (\ref{eq:tkp2}) and (\ref{eq:tkc2}), upper bounds for these constraints can be derived. Consequently, if $(\tilde{\vectw},\tilde{\vect{t}})$ is a feasible point of problem (\ref{eq:genW}) then it holds \cite{AlaaRSCMD} that
\begin{align}\label{DC-approxp}
\frac{|\sbrackets{\vecth_k^{\text{eff}}}^H \vect{\omega}_k^p|^2}
{t_k^p} \geq&
\frac{2\Re\cbrackets{ \sbrackets{\tilde{\vect{\omega}}_k^p}^H \vecth_k^{\text{eff}} \sbrackets{\vecth_k^{\text{eff}}}^H \vect{\omega}_k^p}}
{\overset{}{\tilde{t}_k^p}} -
\frac{|\sbrackets{\vecth_k^{\text{eff}}}^H \overset{\phantom{I}}{\tilde{\vect{\omega}}_k^p}|^2}
{\sbrackets{\tilde{t}_k^p}^2} \, t_k^p,
\end{align}
\begin{align}\label{DC-approxc}
\hspace{0.05cm}\frac{|\sbrackets{\vecth_i^{\text{eff}}}^H \vect{\omega}_k^c|^2}
{t_k^c} \geq&
\frac{2\Re\cbrackets{ \sbrackets{\tilde{\vect{\omega}}_k^c}^H \vecth_i^{\text{eff}} \sbrackets{\vecth_i^{\text{eff}}}^H \vect{\omega}_k^c}}
{\overset{}{\tilde{t}_k^c}} -
\frac{|\sbrackets{\vecth_i^{\text{eff}}}^H \overset{\phantom{I}}{\tilde{\vect{\omega}}_k^c}|^2}
{\sbrackets{\tilde{t}_k^c}^2} \, t_k^c ,
\end{align}
where $\text{Re}\cbrackets{\cdot}$ denotes the real part of a complex-valued number.
By utilizing the approximations (\ref{DC-approxp}) and (\ref{DC-approxc}), inner-convex approximations of the constraints (\ref{eq:tkp2}) and (\ref{eq:tkc2})  can be established by substituting the corresponding terms with their respective upper bound, namely
\begin{align}
      &0 \geq \sum_{j\in\mathcal{K}\text{\textbackslash}\cbrackets{k}} |\sbrackets{\vecth_k^{\text{eff}}}^H \vect{\omega}_j^p|^2 + \sum_{\ell \in \bar{\Phi}_k} |\sbrackets{\vecth_k^{\text{eff}}}^H \vect{\omega}_\ell^c|^2 + \sigma^2-
      \nonumber\\
      &\qquad \qquad\frac{2\Re\cbrackets{ \sbrackets{\tilde{\vect{\omega}}_k^p}^H \vecth_k^{\text{eff}} \sbrackets{\vecth_k^{\text{eff}}}^H \vect{\omega}_k^p}}
      {\overset{}{\tilde{t}_k^p}}+\frac{|\sbrackets{\vecth_k^{\text{eff}}}^H \overset{\phantom{I}}{\tilde{\vect{\omega}}_k^p}|^2}
      {\sbrackets{\tilde{t}_k^p}^2} \, t_k^p, \qquad \forall k \in \mathcal{K},\label{eq:approxCon1}\\
      &0\geq {T_i}+ \sum_{\ell \in \bar{\Phi}_i} |\sbrackets{\vecth_i^{\text{eff}}}^H \vect{\omega}_\ell^c|^2 + \sum_{m \in \Omega_{i,k}} |\sbrackets{\vecth_i^{\text{eff}}}^H \vect{\omega}_m^c|^2-
      \nonumber \\
      &\qquad \qquad\frac{2\Re\cbrackets{ \sbrackets{\tilde{\vect{\omega}}_k^c}^H \vecth_i^{\text{eff}} \sbrackets{\vecth_i^{\text{eff}}}^H \vect{\omega}_k^c}}
      {\overset{}{\tilde{t}_k^c}}+ \frac{|\sbrackets{\vecth_i^{\text{eff}}}^H \overset{\phantom{I}}{\tilde{\vect{\omega}}_k^c}|^2}
      {\sbrackets{\tilde{t}_k^c}^2} \, t_k^c, \qquad\forall k \in \mathcal{K} ,\, \forall i \in \mathcal{M}_k .
     \label{eq:approxCon2}
\end{align}
Next, we focus on finding a convex representation for (\ref{eq:ApproxCap}). To this end, we introduce the slack variables $\vect{q}_k=[q_k^p,q_k^c]^T, \vect{q}=[\vect{q}_1^T,\dots,\vect{q}_K^T]^T, \vect{d}_{k}=[d_{1,k}^p,d_{1,k}^c, \dots, d_{N,k}^p,d_{N,k}^c]^T$ and $\vect{d}=[\vect{d}_1^T,\dots,\vect{d}_K^T]^T$ in order to split the constraint (\ref{eq:ApproxCap}) in problem (\ref{eq:genW}) into five simpler constraints \cite[Proposition 1]{AlaaRSCMD} as follows:
\begin{align}\label{eq:cmpW}
   \underset{\{\vect{w}_k,\vect{t}_k,\mat{R}_k,\vect{d}_{k},\vect{q}_k\}_{k=1}^K}{\text{maximize}}& \quad  \frac{R^{\mathsf{t}}(\mat{R})}{P^{\mathsf{Tr}}(\vectw)+P_\Alpha^{\mathsf{Circ}}(\vectw,\mat{R})} \tag{P2.2}\\
    \text{subject to} \qquad &(\ref{eq:powConsII}), (\ref{eq:QoS}),
   (\ref{eq:p2.1_first})-(\ref{eq:2.1_last}),\,(\ref{eq:approxCon1}),\,(\ref{eq:approxCon2})\nonumber
   \end{align}
   \vspace{-0.95cm}
   \begin{align}
    f_\Alpha\left(||\vectw^p_{n,k}||_2^2\right) &\leq d_{n,k}^p && \forall n \in \mathcal{N},\, \forall k \in \mathcal{K}, \label{eq:cmpW_cite1}\\
    f_\Alpha\left(||\vectw^c_{n,k}||_2^2\right) &\leq d_{n,k}^c && \forall n \in \mathcal{N},\, \forall k \in \mathcal{K}, \label{eq:cmpW_cite2}\\
    \log_2(1+t_k^p) &\leq q_k^p && \forall k \in \mathcal{K},\label{eq:cmpW_log1}\\
    \log_2(1+t_k^c) &\leq q_k^c && \forall k \in \mathcal{K},\label{eq:cmpW_log2}\\
     \sum_{k\in \mathcal{K}} (d_{n,k}^p q_k^p +   d_{n,k}^c q_k^c) &\leq C_n/B,   &&\forall n \in \mathcal{N}. \label{eq:cap_appr_prob}
\end{align}
To solve problem (\ref{eq:cmpW}) this paper adopts \ac{SCA} techniques to transform the problem into convex a representation. Thus, the binilear functions in the constraint (\ref{eq:cap_appr_prob}) can be equivalently rewritten \cite[Proposition 2]{AlaaRSCMD} as
\begin{align}
&(d_{n,k}^p q_k^p +   d_{n,k}^c q_k^c)  \leq  g_{n,k}(\vect{d},\vect{q},\tilde{\vect{d}},\tilde{\vect{q}})\nonumber  = \sum_{o\in\{p,c\}} \big( \frac{1}{2} (d_{n,k}^o + q_k^o)^2  \\  &- \frac{1}{2}(\tilde{d}_{n,k}^o)^2  - (\tilde{q}_{k}^o)^2  - \tilde{d}_{n,k}^o(d_{n,k}^o-\tilde{d}_{n,k}^o) - \tilde{q}_{k}^o(q_{k}^o-\tilde{q}_{k}^o) \big),
\end{align}
where $(\tilde{\vect{d}},\tilde{\vect{q}})$ are feasible fixed values, which satisfy the constraints (\ref{eq:cmpW_cite1})$-$(\ref{eq:cap_appr_prob}). Similarly, equation (\ref{eq:fh_fa}) can be convexified as:
\begin{align}
 \tilde{P}_\Alpha^{\mathsf{FH}}( \vect{d},\vect{q},\tilde{\vect{d}},\tilde{\vect{q}}) =
 P^{\mathsf{Mbps}} \sum_{k\in\mathcal{K}}\sum_{n\in\mathcal{N}}  g_{n,k}( \vect{d},\vect{q},\tilde{\vect{d}},\tilde{\vect{q}} ).
\end{align}
Furthermore, for the constraints (\ref{eq:cmpW_cite1})$-$(\ref{eq:cmpW_cite2}), the concave functions $f_\Alpha(|| \vect{w}_{n,k}^p ||)$ and $f_\Alpha(|| \vect{w}_{n,k}^c||)$ are linearized around $ \tilde{\vect{w}}_{n,k}^p $ and $ \tilde{\vect{w}}_{n,k}^c $, respectively. The resulting convex approximation of the set can be defined by the constraints:
\begin{align}\label{eq:d_approx}
   f_\Alpha(||& \tilde{\vectw^o}_{n,k} ||_2^2)+ \nabla	f_\Alpha(|| \tilde{\vectw^o}_{n,k} ||_2^2)(|| \vectw^o_{n,k} ||_2^2 -  || \tilde{\vectw^o}_{n,k} ||_2^2) \leq d_{n,k}^o ,\forall n\in \mathcal{N}, k\in\mathcal{K}, o \in \{p,c\}.
\end{align}
The same prodecure is applied to the constrains (\ref{eq:cmpW_log1}) and (\ref{eq:cmpW_log2}) by linearizing the concave functions $\log_2(1+t_k^o)$ around $\tilde{t}_k^o$, namely
\begin{align}\label{eq:approx_q}
   \log_2(1+t_k^o)+\frac{1}{(1+\tilde{t}_k^o)\ln(2)}(t_k^o-\tilde{t}_k^o) \leq q_k^o, \qquad \forall k\in \mathcal{K}\,, \forall o\in\{p,c\}.
\end{align}
With the approximations defined above, we are able to formulate an approximate optimization problem as:
\begin{align}\label{eq:prob_w_complete}
&\underset{\{\vect{w}_k,\vect{t}_k,\mat{R}_k,\vect{d}_{k},\vect{q}_k\}_{k=1}^K}{\text{maximize}} \quad  \frac{R^{\mathsf{t}}(\mat{R})}{P^{\mathsf{Tr}}(\vectw)+\tilde{P}_\Alpha^{\mathsf{Circ}}(\vect{d},\vect{q},\tilde{\vect{d}},\tilde{\vect{q}})} = \tilde{E}^{\mathsf{Obj}}(\mat{\Lambda}) \tag{P2.3}\\
    &\text{subject to} \quad  (\ref{eq:powConsII}), (\ref{eq:QoS}),
   (\ref{eq:p2.1_first})-(\ref{eq:2.1_last}),\,(\ref{eq:approxCon1}),\,(\ref{eq:approxCon2}),\,(\ref{eq:d_approx}),\,(\ref{eq:approx_q}),\nonumber\\
    & \qquad \sum_{k\in \mathcal{K}} \left(g_{n,k}(\vect{d},\vect{q},\tilde{\vect{d}},\tilde{\vect{q}})\right) \leq C_n/B, \qquad   \forall n \in \mathcal{N}.
\end{align}
Problem (\ref{eq:prob_w_complete}) is convex and can be solved with the Dinkelbach algorithm \cite{Alaa_EE_RSCMD,dinkelbach}.

Let $\mat{\Lambda} = \hbrackets{\vect{w}^T ,\,\vect{t}^T,\vect{d}^T,\vect{q}^T}^T$ be a vector stacking the optimization variables of (\ref{eq:prob_w_complete}), $\widehat{\mat{\Lambda}}_z = [ \widehat{\vect{w}}_z^T ,\, \widehat{\vect{t}}_z^{\,\,T}\,,\widehat{\vect{d}}_z^T,\widehat{\vect{q}}_z^T]^T $ be the variables that are the optimal solution of problem (\ref{eq:prob_w_complete}) computed at iteration $z$ and $\tilde{\mat{\Lambda}} = [ \tilde{\vect{w}}^T ,\, \tilde{\vect{t}}\phantom{.}^T\, \tilde{\vect{d}}^T,\tilde{\vect{q}}^T]^T $ be the point, around which the approximations are computed. Moreover, let $\mathcal{A}$ be the convex feasible set of problem (\ref{eq:prob_w_complete}).
To solve the problem in an iterative manner, the algorithm starts by initializing vector $\tilde{\mat{\Lambda}}_z$. More specifically, we first initialize the algorithm with feasible \ac{MRT} beamformers $\tilde{\vect{w}}_z$ for the users. Next, $\tilde{\vect{t}}_z$ is initialized with equation (\ref{eq:gammap}). $\tilde{\vect{d}}_z$ and $\tilde{\vect{q}}_z$ are then initialized by replacing the inequalities with equalities in (\ref{eq:cmpW_cite1})$-$(\ref{eq:cmpW_log2}).  %
Using this initialization, problem (\ref{eq:prob_w_complete}) in iteration $z$ can be solved to obtain the vector $\widehat{\mat{\Lambda}}_z$. If the current solution $\widehat{\mat{\Lambda}}_z$ is not stationary, it is used for computing $\tilde{\mat{\Lambda}}_{z+1}$ for the next iteration, i.e., $\tilde{\mat{\Lambda}}_{z+1} = \tilde{\mat{\Lambda}}_z + \varrho_z \sbrackets{\widehat{\mat{\Lambda}}_z - \tilde{\mat{\Lambda}}_z},\,\text{for some }\varrho_z \in (0,1]$, until a stationary solution is found. The detailed steps are outlined in Algorithm \ref{alg:w} below.
%
\begin{algorithm}
\footnotesize
\caption{Procedure to determine the optimal beamforming vector $\vectw^*$ of problem (\ref{eq:genW})}\label{alg:w}
\begin{algorithmic}
\STATE \textbf{Input: } $\tilde{\mat{\Lambda}}_0\in \mathcal{A}$, $Z^{\text{max}} \in \mathbb{N}$
\STATE Initialize: $z\leftarrow 0$, $\widehat{\mat{\Lambda}}_0\leftarrow \tilde{\mat{\Lambda}}_0$
\WHILE{$\widehat{\mat{\Lambda}}_{z}$ is not a stationary solution of problem (\ref{eq:prob_w_complete}) \textbf{and} $z < Z^{\text{max}}$}
   \STATE $z \leftarrow z+1$
   \STATE $\lambda_{z} \leftarrow \tilde{E}^{\mathsf{Obj}}(\widehat{\mat{\Lambda}}_{z-1})$
   \STATE Solve the convex problem (\ref{eq:prob_w_complete}) approximated around $\tilde{\mat{\Lambda}}_{z-1}$ to obtain $\widehat{\mat{\Lambda}}_z$ as:
   \STATE $\widehat{\mat{\Lambda}}_z = \underset{\mat{\Lambda}\in\mathcal{A}}{\text{argmax}} \left\{R^{\mathsf{t}}(\mat{R}) - \lambda_{z}\big({P^{\mathsf{Tr}}(\vectw)+\tilde{P}_\Alpha^{\mathsf{Circ}}(\vect{d},\vect{q},\tilde{\vect{d}}_{z-1},\tilde{\vect{q}}_{z-1}})\big)\right\}$
   \STATE $\tilde{\mat{\Lambda}}_{z} \leftarrow \tilde{\mat{\Lambda}}_{z-1} + \varrho_z \sbrackets{\widehat{\mat{\Lambda}}_{z} - \tilde{\mat{\Lambda}}_{z-1}},\,\text{for some }\varrho_z \in (0,1]$
\ENDWHILE
\STATE \textbf{Output: } $\widehat{\mat{\Lambda}}_z = [ \widehat{\vect{w}}_z^T ,\, \widehat{\vect{t}}_z^{\,\,T}\,,\widehat{\vect{d}}_z^T,\widehat{\vect{q}}_z^T]^T $
\end{algorithmic}
\end{algorithm}

%% file: sections/alternatingMinimization/v_problem.tex
\subsection{Phase Shift Design}
%
Given the vector $\mat{\Lambda} = \hbrackets{\vect{w}^T ,\,\vect{t}^T,\vect{d}^T,\vect{q}^T}^T$, which is assumed to be fixed for the duration of optimizing the phase shift vector $\vectv$, 
Problem (\ref{eq:genProb}) is suitable to be reformulated as a \ac{QCQP} problem {\cite{IRSImprovement1}. To this end, problem (\ref{eq:genProb}) can first be reexpressed as the following feasibility detection problem:
\begin{align}\label{eq:genV-reform1}
   \text{find}\qquad &  \vect{v} \tag{P3.1}\\
   \text{subject to} \qquad
    &\gamma_k^p(\vectv) \geq t_k^p, && \forall k \in \mathcal{K}, \label{eq:reform1_v-p}\\
    &\gamma_{i,k}^c(\vectv)\geq t_k^c,  && \forall k \in \mathcal{K} ,\, \forall i \in \mathcal{M}_k, \label{eq:reform1_v-c}\\
    & | v_r | = 1,  && \forall r \in \{ 1,..., R\},
\end{align}
as only the \ac{SINR} constraints, (\ref{eq:privateRate}) and (\ref{eq:commonRate}), and the unit-modulus constraints are dependent on the phase-shift vector.
However, the feasible set of Problem (\ref{eq:genV-reform1}) is still non-convex. In order to obtain a convex representation of the phase-shift-dependent SINR-constraints (\ref{eq:reform1_v-p}) and (\ref{eq:reform1_v-c}), they can be expressed as
\begin{align}\label{eq:genV-reform2-p}
        &\hspace{-2.25cm}|\left( \vect{h}_k + \mat{H}_k \vect{v} \right)^H \vect{\omega}_k^p|^2 \geq t_k^p \left(
        \sum_{j\in\mathcal{K}\text{\textbackslash}\cbrackets{k}} \hspace{-0.175cm}|\left( \vect{h}_k + \mat{H}_k \vect{v} \right)^H \vect{\omega}_j^p|^2+ \right.\nonumber \\
        & \left. \hspace{2.25cm} \sum_{\ell \in \bar{\Phi}_k} |\left( \vect{h}_k + \mat{H}_k \vect{v} \right)^H \vect{\omega}_\ell^c|^2 + \sigma^2\right), \quad \forall k \in \mathcal{K},
        \end{align}
        \begin{align}
         &\hspace{-0.5cm}|\left( \vect{h}_i + \mat{H}_i \vect{v} \right)^H \vect{\omega}_k^c|^2 \geq t_k^c\sbrackets{
         T_i + \sum_{\ell \in \bar{\Phi}_i} |\left( \vect{h}_i + \mat{H}_i \vect{v} \right)^H \vect{\omega}_\ell^c|^2+ \right. \nonumber \\
         & \left. \hspace{4.75cm} \sum_{m \in \Omega_{i,k}} \hspace{-0.25cm} |\left( \vect{h}_i + \mat{H}_i \vect{v} \right)^H \vect{\omega}_m^c|^2},\quad \forall k \in \mathcal{K} ,\,  \forall i \in \mathcal{M}_k,\label{eq:genV-reform2-c}
\end{align}
with the use of the \ac{SINR} expressions (\ref{eq:gammap}) and (\ref{eq:gammac}).
By denoting
\begin{align}
   b_{k,i}^o &= \vect{h}_k^H \vectw_i^o \text{  and  }
   \vecta_{k,i}^o = \matH_k^H \vectw_i^o, \\
   \mat{M}_{k,j}^o &=
         \begin{bmatrix}
            \vect{a}_{k,j}^o ({\vect{a}_{k,j}^o})^H & (b_{k,j}^o)^H \vect{a}_{k,j}^o \\
            {b_{k,j}^o} ({\vect{a}_{k,j}^o})^H  & 0  \\
         \end{bmatrix},\\
        \tilde{\vect{v}} \,&= \,[\vect{v}^T, s]^T, \label{eq:auxVar}
\end{align}
where $o\in\{p,c\}$ and $s$ is an auxiliary variable, %
it holds that if a feasible solution $\vectvt^*$ is found, the solution $\vect{v}^*$ can be retrieved by
$\vectv^*=[\vectvt^*/\tilde{v}^*_{R+1}]_{(1:R)}$, where $[\vect{x}]_{(1:R)}$ denotes the first $R$ elements of vector $\vect{x}$ and $x_r$ denotes the $r$-th element of vector $\vect{x}$ \cite{IRSImprovement1}. With the above definitions, we are able to formulate a convex representation of the constraints (\ref{eq:genV-reform2-p}) and (\ref{eq:genV-reform2-c}) by utilizing the matrix lifting technique, namely $\matV = \vectvt\vectvt^H$ \cite{federatedLearning}.
This facilitates the reformulation of problem (\ref{eq:genV-reform1}) into
\begin{align}\label{eq:genV-reform4}
  & \text{find}\qquad   \matV \tag{P3.2}\\
  & \text{subject to} \qquad \nonumber \\
    &\sqr{|b_{k,k}^p|} +  \Tr{\matM_{k,k}^p\matV} \geq t_k^p \sbrackets{
        \sum_{j\in\mathcal{K}\text{\textbackslash}\cbrackets{k}} \sqr{|b_{k,j}^p|} +  \Tr{\matM_{k,j}^p\matV}+ \right. \nonumber \\
        & \qquad \left. \qquad\qquad\qquad\qquad\quad \hspace{0.2cm}   \sum_{\ell \in \bar{\Phi}_k} \sqr{|b_{k,\ell}^c|} +  \Tr{\matM_{k,\ell}^c \matV}  + \sigma^2}, \hspace{0.25cm} \forall k \in \mathcal{K} , \label{eq:SINR_p-reform4}\\
    &\sqr{|b_{i,k}^c|}+  \Tr{\matM_{i,k}^c \matV} \geq t_k^c \left(
         \sum_{j \in \mathcal{K}}^{\phantom{I}} \sqr{|b_{i,j}^p|} \hspace{-0.05cm}+ \hspace{-0.05cm} \Tr{\matM_{i,j}^p\matV} \hspace{-0.05cm} + \sigma^2 \hspace{-0.05cm} \right. \left.  + \sum_{\ell \in \bar{\Phi}_i} \sqr{|b_{i,\ell}^c|} + \Tr{\matM_{i,\ell}^c \matV}+ \right. \nonumber \\
         & \qquad \left.\qquad\qquad\qquad\qquad\qquad\quad     \sum_{m \in \Omega_{i,k}} \hspace{-0.18cm} \sqr{|b_{i,m}^c|} + \Tr{\matM_{i,m}^c \matV} \right), \hspace{0.25cm} \forall k \in \mathcal{K} ,\, \forall i \in \mathcal{M}_k , \label{eq:SINR_c-reform4}\\
     &\hspace{3.05cm} V_{r,r}  = 1,  \hspace{6.5cm} \forall r \in \{ 1,..., R+1\},\label{eq:diag}\\
     & \hspace{3.22cm}\matV \succeq 0, \label{eq:psd}\\
     & \hspace{2.12cm}\rank{\matV} =1. \label{eq:genV-reform4_nonConvex}
\end{align}
Problem (\ref{eq:genV-reform4}) is still non-convex due to the rank-one constraint (\ref{eq:genV-reform4_nonConvex}). However, this enables solving the problem with the \ac{SDR} technique, in which the rank-one constraint is dropped in order to obtain a matrix optimization problem, that can be solved with existing solvers. As the obtained solution is not necessarily rank-one, it only acts as an upper-bound of the unrelaxed problem. For this reason, this work proposes the combination of two approaches to find suitable rank-one solutions: 1) The Gaussian randomization technique \cite{GaussRelax} and 2) a \ac{DC} programming approach, that exploits the fact that the nuclear norm and the spectral norm of a \ac{PSD} rank-one matrix have the same values \cite{federatedLearning}.
\subsubsection{Convergence Performance}\label{sec:convBeh}
The shortcomings of utilizing these techniques are that either one does not improve the objective function $E^{\mathsf{Obj}}(\vectw)$ of problem (\ref{eq:genProb}) directly with their respective solutions.
In particular, denoting the objective function $f(\vectw,\vectv)$ based on a feasible solution $(\vectw,\vectv)$, if there exists a feasible solution to (\ref{eq:genW}), and the solution of problem (\ref{eq:genV-reform4}) is obtained by any of the approaches above, i.e., $(\vectw^{(t)},\vectv^{(t+1)})$, where $t$ is the iteration index of the alternating optimization, it holds that
\begin{align}\label{eq:a}
E^{\mathsf{Obj}}(\vectw^{(t)},\vectv^{(t)})= E^{\mathsf{Obj}}(\vectw^{(t)},\vectv^{(t+1)}),
\end{align}
as the objective function $E^{\mathsf{Obj}}(\vectw)$ of (\ref{eq:genW}) is not dependent on the phase shift vector $\vectv^{(t)}$. Furthermore, there is an uncertainty about the objective function increasing after solving the beamforming optimization problem (\ref{eq:genW}) based on the phase shift vector $\vectv^{(t+1)}$, which can be expressed as
\begin{align}\label{eq:b}
E^{\mathsf{Obj}}(\vectw^{(t)},\vectv^{(t+1)}) \nleq E^{\mathsf{Obj}}(\vectw^{(t+1)},\vectv^{(t+1)}),
\end{align}
because the retrieved solution $\vectv^*$ might not satisfy the constraints of problem (\ref{eq:genV-reform1}) due to the refomulation into a \ac{QCQP}.
This results in the following expression
\begin{align}\label{eq:ab}
 E^{\mathsf{Obj}}(\vectw^{(t)},\vectv^{(t)}) &\overset{(\ref{eq:a})}{=} E^{\mathsf{Obj}}(\vectw^{(t)},\vectv^{(t+1)})  \overset{(\ref{eq:b})}{\nleq} E^{\mathsf{Obj}}(\vectw^{(t+1)},\vectv^{(t+1)}),
\end{align}
which captures the change of the objective function $E^{\mathsf{Obj}}$ during the alternating optimization. From (\ref{eq:ab}) it can be inferred that both approaches in combination with the alternating optimization framework are not guaranteed to converge to a locally optimal solution \cite{rankOneRewrite}. More precisely, the objective values obtained from problem (\ref{eq:genW}) before $ \left( E^{\mathsf{Obj}}(\vectw^{(t)},\vectv^{(t)})\right)$ and after $\left( E^{\mathsf{Obj}}(\vectw^{(t+1)},\vectv^{(t+1)}) \right)$, updating the phase shift vector $\vectv$ are determined on two different sets of phase shift vectors, namely $\vectv^{(t)}$ and $\vectv^{(t+1)}$, respectively.
Hence, updating the phase shift vectors from $\vectv^{(t)}$ to $\vectv^{(t+1)}$ also results in a change of the feasible set of problem (\ref{eq:genV-reform4}).
Consequently, there is no guarantee of the optimal objective value of problem (\ref{eq:genW}) improving  after updating the phase shift vector from $\vectv^{(t)}$ to $\vectv^{(t+1)}$ because the relation between the two underlying feasible sets can not be quantified as both, $\vectv^{(t)}$ and $\vectv^{(t+1)}$, may be infeasible for problem (\ref{eq:genV-reform1}), due to the auxiliary variable $s$ and the use of the \ac{SDR} technique.

For these reasons, this work proposes to find phase shifters, which exclusively increase the private and common \acp{SINR} at each user. In order to still guarantee a feasible rank-one solution, this approach is combined with the Gaussian randomization technique and the \ac{DC} programming approach. The combination of these approaches, i.e., the hybrid approach, facilitates the reformulation of problem (\ref{eq:genV-reform4}) into an optimization problem that not only guarantees the convergence of the alternating optimization but also obtains phase shifters, which are more efficient towards maximizing the energy efficiency of the network. This results in expression (\ref{eq:b}) becoming
\begin{align}\label{eq:tildeB}
E^{\mathsf{Obj}}(\vectw^{(t)},\vectv^{(t+1)}) \leq E^{\mathsf{Obj}}(\vectw^{(t+1)},\vectv^{(t+1)}),
\end{align}
since it is assured that the objective value does not decrease after updating the phase shifters $\vectv$.


In order to satisfy (\ref{eq:tildeB}), the next section utilizes this hybrid approach. To this end, 
slack variables are defined, which are maximized in order to obtain feasible solutions for the phase shift vector, while supporting the objective of maximizing the energy efficiency.
\subsubsection{Hybrid Approach}
With the introduction of the slack variables $\zeta^p_k$ and $\zeta_{k}^p$, problem (\ref{eq:genV-reform4}) can be extended to
\begin{align}\label{eq:genV-alpha}
&\underset{\matV,\{\zeta_k^p,\zeta_{k}^c\}_{k=1}^K}{\text{maximize}}\qquad \sum_{k \in \mathcal{K}}\zeta_k^p + \zeta_{k}^c  \tag{P3.3}\\
   &\text{subject to} \qquad
    \sqr{|b_{k,k}^p|} +  \Tr{\matM_{k,k}^p\matV} \geq \eta  t_k^p  \Big(
        \sum_{j\in\mathcal{K}\text{\textbackslash}\cbrackets{k}} \sqr{|b_{k,j}^p|} +  \Tr{\matM_{k,j}^p\matV} +
          \sum_{\ell \in \bar{\Phi}_k}  \sqr{|b_{k,\ell}^c|}   \nonumber \\ & \qquad \qquad +\Tr{\matM_{k,\ell}^c \matV}  + \sigma^2 \Big) + \zeta_k^p,  \hspace{0.5cm} \forall k \in \mathcal{K}  \label{eq:SINR_p-reform5}\\
    &\sqr{|b_{i,k}^c|}+  \Tr{\matM_{i,k}^c \matV} \geq \eta t_k^c \Big(
         \sum_{j \in \mathcal{K}}^{\phantom{I}} \sqr{|b_{i,j}^p|} +  \Tr{\matM_{i,j}^p\matV} + \sigma^2   + \sum_{\ell \in \bar{\Phi}_i} \sqr{|b_{i,\ell}^c|} + \Tr{\matM_{i,\ell}^c \matV} \nonumber \\ &  \qquad\qquad+ \sum_{m \in \Omega_{i,k}} \sqr{|b_{i,m}^c|} +   \Tr{\matM_{i,m}^c \matV} \Big) + \zeta_{k}^c,  \hspace{0.35cm} \forall k \in \mathcal{K} ,\, \forall i \in \mathcal{M}_k  \label{eq:SINR_c-reform5},\\
    &\qquad\qquad\qquad(\ref{eq:diag}),(\ref{eq:psd}),(\ref{eq:genV-reform4_nonConvex}),\nonumber
\end{align}
where $\zeta^p_k$ and $\zeta_{k}^p$ can be understood as \textit{SINR residuals} of user $k$ in phase shift optimization \cite{IRSImprovement1} and $\eta \in [0,1]$ is a parameter, which regulates the objective of the phase shift optimization. In fact, for $\eta=0$ the resulting phase shifters maximize the sum-path gain, which ignores the users' interference-and-noise term of the private and common \acp{SINR} and therefore comes at the cost of increased interference within the network. The rationale behind this approach is to maximize the potential gains of the phase shifters on the network, while subsequently utilizing the dynamic \ac{CMD} set allocation to cancel the resulting increase in interference. The other extreme case, namely $\eta=1$, changes the objective of problem (\ref{eq:genV-alpha}) to find phase shifters that strictly increase the private and common \ac{SINR}s at the users, which ensures convergence as (\ref{eq:tildeB}) holds. This approach also aids in increasing the \ac{EE} of the network because increasing the rate at the users results in a higher total rate of the network $R^t$. Thus, it also enables the beamforming optimization to scale the beamformers $\vectw$ down in case the fronthaul is already at full capacity.

Next, we focus on reformulating the rank-one constraint (\ref{eq:genV-reform4_nonConvex}) into the equivalent \ac{DC} form. To this end, (\ref{eq:genV-reform4_nonConvex}) can be reformulated using the following proposition:
\begin{prop}\label{prop:1}
 \textit{For a \ac{PSD} matrix $\mat{X} \in \mathbb{C}^{N\times N}$ and }$\myNorm{\mat{X}}_* \geq 0$\textit{, it holds \cite{federatedLearning} that}
\end{prop}
\begin{align}\label{eq:eqR1}
   \text{rank}(\mat{X}) = 1 \Leftrightarrow \myNorm{\mat{X}}_* - \myNorm{\mat{X}}_2 = 0,
\end{align}
\textit{where $\myNorm{\mat{X}}_*$ denotes the nuclear norm and $\myNorm{\mat{X}}_2$ denotes the spectral norm of $\mat{X}$.}
To tackle the \ac{DC} form of the reformulated rank-one expression, the first-order Taylor approximation of the spectral norm $\myNorm{\mat{V}}_2$ around point $\mat{V}^0$ can be derived as
\begin{align}\label{convCons1}
   \myNorm{\mat{V}}_2 \geq \myNorm{\mat{V}^0}_2 + \langle \partial_{\mat{V}^0} \myNorm{\mat{V}}_2 , \sbrackets{\mat{V}-\mat{V}^0}\rangle,
\end{align}
where $\partial_{\mat{V}^0}$ is the subgradient of $\myNorm{\mat{V}}_2$ with respect to $\matV$ at $\matV^0$ and where the inner product is defined as
$\langle \mat{X},\mat{Y} \rangle = \text{Re}\cbrackets{\Tr{\mat{X}^H\mat{Y}}}$
as stated by Wirtinger's calculus \cite{wirtingers} in the complex domain. Using these definitions, the spectral norm in (\ref{eq:eqR1}) can be replaced with the approximation given in (\ref{convCons1}) to obtain a convex approximation of the rank-one constraint. The resulting expression is suitable to be added as 
a penalty term to problem (\ref{eq:genV-alpha}) and results in the following optimization problem
\begin{align}\label{eq:genV-alpha_rank}
   &\underset{\matV,\{\zeta_k^p,\zeta_{k}^c\}_{k=1}^K}{\text{maximize}} \quad \rho \Big( \sum_{k \in \mathcal{K}}  \hspace{-0.05cm} \zeta_k^p + \zeta_{k}^c \Big) - \,  (1-\rho)  \tag{P3.4}\Big(\hspace{-0.05cm}\myNorm{\mat{V}}_* - \myNorm{\mat{V}^0}_2 - \langle \partial_{\mat{V}^0} \myNorm{\mat{V}}_2 , \sbrackets{\mat{V}-\mat{V}^0}\rangle\hspace{-0.05cm} \Big) \nonumber \\
   &\text{subject to} \hspace{0.5cm}  (\ref{eq:diag}), (\ref{eq:psd}), (\ref{eq:SINR_p-reform5}), (\ref{eq:SINR_c-reform5}), \nonumber
\end{align}
where $\rho \in [0,1]$ regulates the trade-off between a high-quality and a rank-one solution. 
It is worth noting that the subgradient $\partial_{\mat{V}^0}\myNorm{\mat{V}}_2$ can be efficiently computed by using the following proposition \cite{federatedLearning}.
\begin{prop}\label{prop:2}
\textit{For a given \ac{PSD} matrix $\mat{X}\in\mathbb{C}^{N\times N} $, the subgradient  $\partial_{\mat{X}}\myNorm{\mat{X}}_2$ can be computed as $\vect{e}_1\vect{e}_1^H$, where $\vect{e}_1 \in \mathbb{C}^N$ is the leading eigenvector of matrix $\mat{X}$.}
\end{prop}
Problem (\ref{eq:genV-alpha_rank}) is convex and solvable with existing solvers. After obtaining the solution $\matV^*$ of problem (\ref{eq:genV-alpha_rank}), the Gaussian randomization technique is applied to determine a feasible solution of high quality to problem (\ref{eq:genV-reform1}). To this end, the \ac{SVD} of $\matV^*$  is calculated  as
$\matV^* = \mat{U} \mat{\Sigma} \mat{U}^H$,
where $\mat{U} \in \mathbb{C}^{(R+1)\times(R+1)}$ and $\mat{\Sigma} \in \mathbb{C}^{(R+1)\times(R+1)}$ are the unitary matrix and the diagonal matrix, respectively. A rank-one candidate solution $\hat{\vectv}_g$ to problem (\ref{eq:genV-reform1}) can be generated \cite{GaussRandHelp} with the use of the \ac{SVD} components, namely
\begin{align}\label{eq:candidate}
   \hat{\vectv}_g = \mat{U}\mat{\Sigma}^{\frac{1}{2}} \vect{z}_g \in \mathbb{C}^{(R+1)\times 1},
\end{align}
where $\vect{z}_g \sim \mathcal{CN}(\zero_{R+1},\mat{I}_{R+1})$ denotes a random vector drawn independently from a circularly-symmetric complex Gaussian distribution. After $G$ randomized solutions are generated, each randomization $\hat{\vectv}_g$ can be used to obtain a potential candidate solution $\overline{\vectv}_g$ for problem (\ref{eq:genV-reform1}), namely
\begin{align}\label{eq:finalVsol}
   \displaystyle \overline{\vectv}_g = \exp\left({\displaystyle j \arg{ \sbrackets{ \hbrackets{{\hat{\vectv}}/{\hat{v}_{R+1}}}_{(1:R)}}  }}\right).
\end{align}
If a potential candidate solution $\overline{\vectv}_g$ satisfies the constraints of problem (\ref{eq:genV-reform1}), it is declared to be a feasible candidate solution. The best performing solution out of all feasible candidates, i.e., the one with the highest achievable total rate of the network, is chosen as the final solution $\vectv^*$ to problem (\ref{eq:genV-reform1}). If no feasible solution can be determined with the given amount of Gaussian randomizations $G$ and $\eta \neq 1$, the vector $\vect{u}\in \mathbb{C}^{(R+1)\times1}$ associated with the leading singular value is adapted by applying equation (\ref{eq:finalVsol}) and chosen as the solution $\vectv^*$ instead.
Algorithm \ref{alg:v} summarizes the necessary steps towards the goal of obtaining a solution to problem (\ref{eq:genV-reform1}).

\begin{algorithm}
\footnotesize
\caption{Procedure to determine the optimal phase shift vector $\vectv^*$ of problem (\ref{eq:genV-reform1})}\label{alg:v}
\begin{tikzpicture}[>={Latex[length=7pt]}]\label{alg:v}
   \tikzset{myRect/.style={draw,rectangle,minimum width=1.5cm, minimum height=1cm,align=center}}
   \tikzset{myDiam/.style={draw,diamond,aspect=2.25,minimum width=1.5cm, minimum height=1cm,align=center}}
   \tikzset{myArrow/.style={->,draw,line width=0.25mm}}


   \node at (0,1.1){Input: $\{\widehat{\mat{\Lambda}}_0$, $\vectv_0\}$};

   \node[myRect] (source) at (0,0){$g_\text{max}\leftarrow 0,\,R_{\mathsf{max}}^{\mathsf{t}}\leftarrow 0, $ \\$
   \vectvt\leftarrow[\vect{v},1]^T, \, \matV^0 \leftarrow        \vectvt\vectvt^H$};

   \node[myRect] (subg) at ($(source)+(3.7,0)$){Compute $\partial_{\mat{V}^0}\myNorm{\mat{V}}_2$\\(e.g. Proposition \ref{prop:2})};

   \node[myRect] (V_opt) at ($(subg)+(2.9,0)$){Solve (\ref{eq:genV-alpha_rank})};
   \node[myRect] (SVD) at ($(V_opt)+(3.1,0)$){$\matV^* \overset{SVD}{=} \mat{U} \mat{\Sigma} \mat{U}^H$};

   \node[myDiam] (g_leq_G) at ($(SVD)+(0,-1.7)$){};
   \node at (g_leq_G){$g<G$};

   \node[myDiam,minimum width=2.6cm, minimum height=1.5cm] (gmax_diam) at ($(g_leq_G)+(2.6,0)$){};

   \node[align=center] at ($(gmax_diam)+(0,0.025)$){$g_\text{max}=0$\\[-2pt] \textbf{and} $\eta\neq1$};

   \node[myRect] (vg) at ($(g_leq_G)-(2.4,0)$){Generate $\vect{z}_g$,\\
   $\hat{\vectv}_g$ with (\ref{eq:candidate}),\\
   $\overline{\vectv}_g$ with (\ref{eq:finalVsol})};

   \node[myDiam,minimum width=2.95cm, minimum height=1.7cm] (satisfy) at ($(vg)-(2.9,0)$){};
   \node[align=center] at ($(satisfy)+(0,0.025)$) {$\overline{\vectv}_g$ satisfies\\(\ref{eq:reform1_v-p}) and (\ref{eq:reform1_v-c})};

   \node[myDiam,minimum width=2.7cm, minimum height=1.5cm] (Rmax) at ($(satisfy)-(3.25,0)$){};
   \node[align=center] at ($(Rmax)+(0,0.015)$){$R_{g}^\mathsf{t}(\overline{\vectv}_g)>R_{\text{max}}^\mathsf{t}$};

   \node(sup-dot_inv) at ($(Rmax)+(-1.5,-1.5)$){};

   \node[circle,fill=black,draw,scale=0.35](sup-dot1) at ($(Rmax)+(0,-1.5)$){};
   \node[circle,fill=black,draw,scale=0.35](sup-dot2) at ($(satisfy)+(0,-1.5)$){};

%
%
%
%
%
%
%
%
   \draw[myArrow] (0,0.9) -- (source);

\draw[myArrow] ($(source.0)+(0,0)$) -- ($(subg.180)+(0,0)$);
\draw[myArrow] ($(subg.0)+(0,0)$) -- ($(V_opt.180)+(0,0)$);
\draw[myArrow] ($(V_opt.0)+(0,0)$) --node[above,pos=0.45]{$\matV^*$} ($(SVD.180)+(0,0)$);

\draw[myArrow] ($(SVD.270)+(0,0)$) --node[right,pos=0.4]{$g\leftarrow 0$}node[left,pos=0.4]{$\vect{u}$} ($(g_leq_G.90)+(0,0)$);

\draw[myArrow] (g_leq_G) --node[above,pos=0.4]{F} (gmax_diam);
\draw[myArrow] (g_leq_G) --node[above,pos=0.4]{T} (vg);
\draw[myArrow] (vg) -- (satisfy);
\draw[myArrow] (satisfy) --node[above,pos=0.4]{T} (Rmax);

\draw[myArrow] (Rmax) --node[right,pos=0.35]{F} (sup-dot1);
\draw[myArrow] (satisfy) --node[right,pos=0.35]{F} (sup-dot2);

\draw[myArrow,align=right] (Rmax) -|node[above,pos=0.15]{T}node[left,pos=0.78]{$\overset{\phantom{\_}}{R_{\mathsf{max}}^\mathsf{t}}\leftarrow R_{g}^\mathsf{t}$\\$\underset{\phantom{-}}{g_\text{max}} \leftarrow g$} (sup-dot_inv.center) -- (sup-dot1);

\draw[myArrow] (sup-dot1) -- (sup-dot2);
\draw[myArrow] (sup-dot2) -|node[pos=0.7,left]{$g\leftarrow g+1$} (g_leq_G);

\draw[myArrow] (gmax_diam) --node[pos=0.45,right]{F} ($(gmax_diam)+(0,1.2)$);

\draw[myArrow] (gmax_diam) --node[pos=0.45,right]{T} ($(gmax_diam)-(0,1.2)$);

\node[align=center] at ($(gmax_diam)+(0,1.65)$){Output:\\$\vectv^* \leftarrow \overline{\vectv}_{g_\text{max}}$};

\node[align=left] at ($(gmax_diam)-(0,1.5)$){Output: \\ $ \vect{v}^* \leftarrow \e^{{ j \arg{ \sbrackets{ \hbrackets{{{\vect{u}}}/{{u}_{R+1}}}_{(1:R)}}  }}}$};

%
%
%
%
%
%
%
\end{tikzpicture}
\end{algorithm}

%% file: sections/alternatingMinimization/cmd_problem.tex
\subsection{Determining the CMD sets}
In this work, the \ac{RS} technique is adopted to mitigate the increased impact of the interference in the network, which is caused by the spatial reuse technique and the \ac{IRS}. Determining the optimal \ac{CMD} sets for the scenario at hand is vital for the performance of \ac{RS} because the common rate $R_k^c$ of user $k$ is chosen such that its decodable at all users in the corresponding set $\mathcal{M}_k$. To elaborate, suppose that there is a user $j$ in the group $\mathcal{M}_k$ (users decoding the common message of user $k$) and the user's SINR $\gamma_{j,k}$ of receiving the common message of user $k$ is notably lower than for the rest of the group, the resulting transmission of the common message $s_k^c$  becomes inefficient. This is because the common rate $R_k^c$ of the group $\mathcal{M}_k$ decreases to the rate user $j$ is able to decode, and subsequently reduces the impact of the \ac{SIC} at the users. Specifically for \ac{IRS}-assisted networks, finding suitable \ac{CMD} sets becomes a problem as current methods for determining the \ac{CMD} sets, e.g., with \ac{MRC} beamformers \cite{AlaaRSCMD}, would depend on the initial (sub-optimal) \ac{IRS} phase shifters. Since the phase shifters are subject to change during the optimization process, it becomes inefficient to determine the \ac{CMD} groups based on the initial phase shifters. Moreover, the optimization of the phase shifters in problem (\ref{eq:genV-alpha_rank}) is also dependent on the \ac{CMD} sets in constraint (\ref{eq:SINR_c-reform5}), which increases the difficulty of finding a good combination of effective \ac{CMD} sets and a phase shift vector that efficiently supports the private and common transmissions of the network.

To bridge this mutual dependency of the phase shift vector optimization and the allocation of {CMD} sets, we propose the following method: First, we prioritize the maximization of the achievable sum-path gain in the first few phase shift optimizations, which comes at the cost of potentially increasing the interference at some users. After each phase-shift optimization, efficient \ac{CMD} groups are formed, based on the received interference at the users and potential achievable common rates within the groups. We implement this idea, by initialing the alternating optimization algorithm with a low-value for the parameter $\eta \in [0,1]$, which increases for each successive iteration of the phase shift optimization, so that (\ref{eq:tildeB}) holds and convergence can be assured.

To this end, let $\tilde{\gamma}_{i,k}^p$ denote the \ac{SINR} of user $i$ decoding the \textit{private} message of user $k$ as:
\begin{align}
   \tilde{\gamma}_{i,k}^p(\vectw,\vectv) &= \label{eq:gammap_diff}\frac{|{\sbrackets{\vecth_i^{\text{eff}}(\vectv)}^H \vectw_k^p}|^2}
                     {\sum\limits_{m \in \mathcal{K}\text{\textbackslash}\cbrackets{k}} \hspace{-0.2cm} |{\sbrackets{\vecth_i^{\text{eff}}(\vectv)}^H \vectw_m^p}|^2 + \sum\limits_{\ell \in \bar{\Phi}_k} |{\sbrackets{\vecth_i^{\text{eff}}(\vectv)}^H \vectw_\ell^c}|^2 + \sigma^2}.
\end{align}
Moreover, let $\vect{\Gamma}_k^o, \,\forall o \in \{p,c\}$ denote a vector, whose entries represent the ratio of each user decoding the private/common message of user $k$ and user $k$ decoding his private/common message:
\begin{align}\label{eq:cmdGam_p}
\vect{\Gamma}_k^p = [\frac{\tilde{\gamma}_{1,k}^p(\vectw,\vectv)}{\gamma_k^p(\vectw,\vectv)}-1 , \dots , \frac{\tilde{\gamma}_{K,k}^p(\vectw,\vectv)}{\gamma_k^p(\vectw,\vectv)}-1]^T, \\
\vect{\Gamma}_k^c = [\frac{{\gamma}_{1,k}^c(\vectw,\vectv)}{\gamma_{k,k}^c(\vectw,\vectv)}-1 , \dots , \frac{{\gamma}_{K,k}^c(\vectw,\vectv)}{\gamma_{k,k}^c(\vectw,\vectv)}-1]^T, \label{eq:cmdGam_c}
\end{align}
and let $\mat{\Gamma}^o = [\vect{\Gamma}_1^o,\dots,\vect{\Gamma}_K^o]$.
Note that each positive non-zero value in $\vect{\Gamma}_k^o$ represents a user, which is able to decode the respective message of user $k$ with a higher rate than user $k$ itself. The algorithm starts by calculating $\mat{\Gamma}^o$ given the beamforming vectors $\vectw$ and phase shift vector $\vectv$. To ignore the users $j\in\mathcal{M}_k$, which are already decoding the common message of user $k$, the corresponding $(j,k)$-th entries in $\mat{\Gamma}^o$ are replaced with an arbitrarily low value, i.e., $-\infty$. Next, the highest value $\Gamma_{j,k}^o$ of $\mat{\Gamma}^o$ is determined. Should this $(j,k)$-th entry $\Gamma_{j,k}^o$ in $\mat{\Gamma}^o$ be above $\epsilon^\mathsf{CMD} \in [-1,\infty]$ then user $j$ is considered as an addition to the group $\mathcal{M}_k$, if the number of decoding layers $D$ of user $j$ is not at full capacity, i.e., $|\Phi_j| < D$. The effectiveness of this procedure is verified by temporarily adding the user to the respective CMD sets in $\mathcal{S}$, resulting in $\tilde{\mathcal{S}}$. Should the $(j,k)$-th entry $\Gamma_{j,k}^o$ be part of $\mat{\Gamma}^p$, i.e., $o \in \{p\}$, the beamforming vector $\vectw_k^c$, which is associated with the common message of user $k$, is set as $\vectw_k^c = \vectw_k^c + \vectw_k^p$. Should the resulting value $({{\gamma}_{1,k}^c(\vectw,\vectv)}/{\gamma_{k,k}^c(\vectw,\vectv)})-1$, which is determined with the temporary CMD sets $\tilde{\mathcal{S}}$, still be above a certain threshold $\epsilon^\mathsf{Thr} \in [-1,\infty]$ the temporary CMD sets $\tilde{\mathcal{S}}$ are adopted and the corresponding values in $\mat{\Lambda}$ updated. Otherwise the changes to $\vectw_k^c$ are discarded and the corresponding entry of $\mat{\Gamma}^o$ is set to $-\infty$. This process is repeated until there are no values in $\mat{\Gamma}^o$ above $\epsilon^\mathsf{CMD}$. The detailed steps for this procedure are outlined in Algorithm \ref{alg:cmdsets}.

\begin{algorithm}
\footnotesize
\caption{Procedure to dynamically allocate \ac{CMD} sets $\mathcal{S}$}
\begin{tikzpicture}[>={Latex[length=7pt]}]\label{alg:cmdsets}
   \tikzset{myRect/.style={draw,rectangle,minimum width=1.5cm, minimum height=1cm,align=center}}
   \tikzset{myDiam/.style={draw,diamond,aspect=2.25,minimum width=1.5cm, minimum height=1cm,align=center}}
   \tikzset{myArrow/.style={->,draw,line width=0.25mm}}


   \node at (0,1.1){Input: $\{\mathcal{S}$, $\mat{\Lambda}$, $\vectv\}$};

   \node[myRect] (source) at (0,0){Calculate $\mat{\Gamma}^p$ and $\mat{\Gamma}^c$ \\ with (\ref{eq:cmdGam_p}) and (\ref{eq:cmdGam_c}) };

   \node[myRect] (init) at ($(source)+(0,-2.25)$) { Set $(j,k)$-th entry in $\mat{\Gamma}^o$\\ as $\Gamma_{j,k}^o \leftarrow -\infty$};

   \node[myDiam,minimum width=2.75cm, minimum height=1.15cm](diam_cmd) at ($(init)+(0,-1.5)$){};
   \node at (diam_cmd){ $\exists \Gamma_{j,k}^o > \epsilon^\mathsf{CMD}$};

   \node[myRect] (mu) at ($(diam_cmd)+(3.5,0)$) {$\mu_{j,k}^o \leftarrow \underset{o\,\in\{p,c\}}{\max}\{\mat{\Gamma}^o\} $};

   \node[myDiam] (Phi_check) at ($(mu)+(4,0)$) {$|\Phi_j| < D$};

   \node[myRect] (Phi_add) at ($(Phi_check)+(3,0)$) {$\tilde{\Phi}_j \leftarrow \Phi_j \cup \{k\}$};

   \node[myRect] (S) at ($(Phi_add)+(2.75,0)$) {Determine $\tilde{\mathcal{S}}$};

   \node[myDiam] (diam_o) at ($(S)+(0,1.4)$) {$o = p$};

   \node[myRect] (w_p_+) at ($(diam_o)+(0,1.6)$) {$\vect{w}_k^c\leftarrow \vect{w}_k^c+\vect{w}_k^p$};

   \node[myDiam,minimum width=4cm, minimum height=2.25cm] (thresh_diam) at ($(w_p_+)+(-4,0)$) {};

   \node at ($(thresh_diam)+(0.1,0)$) {$\displaystyle\frac{{\gamma}_{1,k}^c(\vectw,\vectv)}{\gamma_{k,k}^c(\vectw,\vectv)}-1 > \epsilon^{\mathsf{Thr}}$};

   \node[myRect] (S_true) at ($(thresh_diam)+(-3.5,0)$) {$\mathcal{S}\leftarrow \tilde{\mathcal{S}}$ and \\ update $\mat{\Lambda}$};

   \node[myRect] (w_p_-) at ($(S_true)+(0,1.5)$) {$\vect{w}_k^c\leftarrow \vect{w}_k^c-\vect{w}_k^p$};

   \node[circle,fill=black,draw,scale=0.35](sup-dot) at ($(init)+(3.5,0)$){};


   \draw[myArrow] (source) -- node[pos=0.5,anchor=180,align=left]{ $\forall j\in\mathcal{M}_k$ \\ $\forall k\in\mathcal{K}$}node[pos=0.5,anchor=0,align=right]{$\forall o\in\{p,c\}$} (init);

   \draw[myArrow] (init) -- (diam_cmd);

   \draw[myArrow] (diam_cmd) --node[pos=0.4,above]{T} (mu);

   \draw[myArrow] (diam_cmd) --node[pos=0.4,right]{F} ($(diam_cmd)+(0,-1)$);
   \node at ($(diam_cmd)+(0,-1.2)$){Output: $\{\mathcal{S}^* \leftarrow \mathcal{S},\mat{\Lambda}\}$};

   \draw[myArrow] (mu) --node[pos=0.45,above]{$(j,k,o)$} (Phi_check);

   \draw[myArrow] (Phi_check) --node[pos=0.38,above]{T} (Phi_add);

   \draw[myArrow] (Phi_add) -- (S);

   \draw[myArrow] (S) -- (diam_o);
   \draw[myArrow] (diam_o) --node[pos=0.2,right]{T} (w_p_+);
   \draw[myArrow] (w_p_+) -- (thresh_diam);
   \draw[myArrow] (diam_o) -|node[pos=0.025,above]{F} (thresh_diam);

   \draw[myArrow] (thresh_diam) --node[pos=0.35,above]{T} (S_true);
   \draw[myArrow] (thresh_diam) |-node[pos=0.2,right]{F} (w_p_-);

   \draw[myArrow] (Phi_check) |- node[pos=0.25,right]{F} ($(Phi_check)!0.6!(sup-dot)$) -|  (sup-dot);

   \draw[myArrow] (S_true) -| ($(S_true)!0.5!(sup-dot)$) |-  (sup-dot);

   \draw[myArrow] (w_p_-) -| (sup-dot);

   \draw[myArrow] (sup-dot) --node[pos=0.45,above]{$(j,k,o)$} (init);

   \draw[myArrow] (0,0.9) -- (source);

\end{tikzpicture}
\end{algorithm}

%% file: sections/alternatingMinimization/obtaining_optimal_solution.tex
The proposed alternating optimization algorithm for solving problem (\ref{eq:genProb}) is outlined in Algorithm \ref{alg:final}, where problems (\ref{eq:prob_w_complete}) and (\ref{eq:genV-alpha_rank}) are solved alternatively, while the objective of the first few outer-loop iterations is to find suitable \ac{CMD} sets by prioritizing the sum-path gain in the phase shift optimization.
\begin{algorithm}
\footnotesize
\caption{Procedure to determine $\mat{\Lambda}^*$, $\vectv^*$ and ${\mathcal{S}^*}$ of problem (\ref{eq:genProb})}
\begin{tikzpicture}[>={Latex[length=7pt]}]\label{alg:final}
   \tikzset{myRect/.style={draw,rectangle,minimum width=1.5cm, minimum height=1cm,align=center}}
   \tikzset{myDiam/.style={draw,diamond,aspect=2.25,minimum width=1.5cm, minimum height=1cm,align=center}}
   \tikzset{myArrow/.style={->,draw,line width=0.25mm}}


   \node at (0,1.1){Input: $\{\widehat{\mat{\Lambda}}_0$, $\vectv_0\}$};

   \node[myRect] (source) at (0,0){$\big\{\Phi_k=\{k\}\big\}_{k\in\mathcal{K}}$ \\ and determine $\overset{\phantom{.}}{\mathcal{S}_0}$};
   
   \node[myRect] (beamf) at ($(source)+(3.7,0)$){Solve (\ref{eq:genW}) \\ (Algorithm \ref{alg:w})};

   \node[myRect] (v_vec) at ($(beamf)+(3.3,0)$){Solve (\ref{eq:genV-reform1})\\ (Algorithm \ref{alg:v})};

   \node[myDiam,minimum width=2.2cm, minimum height=1.22cm] (v_valid) at ($(v_vec)+(3.2,0)$){};
   \node at (v_valid){Valid $\vectv_{z+1}$?};
   
   \node[myRect] (S) at ($(v_valid)+(0,-1.5)$){Determine $\mathcal{S}_{z+1}$\\ (Algorithm \ref{alg:cmdsets})};
   
   \node[myRect] (output) at ($(S)+(2.25,0)$){${\mat{\Lambda}}^* \leftarrow \widehat{\mat{\Lambda}}_{z}$ \\$\vectv^* \leftarrow \vectv_{z}$ \\ $S^*\leftarrow S$};
   
   \node[myDiam,minimum width=2.8cm, minimum height=1.3cm] (Obj_diam) at ($(S)+(0,-1.5)$){};
   \node at (Obj_diam){$E^{\mathsf{Obj}}$ converged?};
   
   \node[myRect] (eta) at ($(Obj_diam)-(4,0)$){$\eta\leftarrow \min\{\eta+0.25,1\}$};

   \draw[myArrow] (0,0.9) -- (source); 

   \draw[myArrow] ($(source.0)+(0,0)$) --node[above,pos=0.45]{$z\leftarrow0$}node[below,pos=0.45]{$\eta\leftarrow0$} ($(beamf.180)+(0,0)$);
   
   \draw[myArrow] ($(beamf.0)+(0,-0)$) --node[above,pos=0.45]{$\widehat{\mat{\Lambda}}_{z+1}$} ($(v_vec.180)+(0,-0)$);
   
   \draw[myArrow] ($(v_vec.0)+(0,0)$) --node[above,pos=0.45]{$\vectv_{z+1}$} ($(v_valid.180)+(0,0)$);
   
   \draw[myArrow] (v_valid) --node[right,pos=0.45]{T} (S);
   
   \draw[myArrow] (S) -- (Obj_diam);
   
   \draw[myArrow] (v_valid) -| node[right,pos=0.1,above]{F}node[right,pos=0.65]{$\mathcal{S}\leftarrow\mathcal{S}_z$} (output);
   \draw[myArrow] (Obj_diam) -| node[right,pos=0.1,above]{T}node[right,pos=0.65]{$\mathcal{S}\leftarrow\mathcal{S}_{z+1}$} (output);
   
   \draw[myArrow] (output) --node[right,pos=0.7,anchor=180,align=center]{Output:\\  $\{{\mat{\Lambda}}^*$,$\vectv^*$,$\mathcal{S}\}$ } ($(output)+(1.25,0)$);
   
   \draw[myArrow] (Obj_diam) --node[right,pos=0.1,above]{F} (eta);
   
   \draw[myArrow]  (eta) -|node[right,pos=0.8]{$z\leftarrow z+1$} (beamf);
\end{tikzpicture}
\end{algorithm}

%% file: sections/convergence_behaviour.tex

\section{Numerical Simulations}\label{ch:numsim}

For the simulation results we consider a \ac{C-RAN} consisting of one \ac{CP}, which is connected to 3 \acp{BS} via finite capacity fronthaul links. Each \ac{BS} is equipped with 2 antennas and serving 6 single-antenna users in the area of operation, sized as $[-500,500] \times [-500,500] \text{ m}^2 $. It is assumed that all users require a \ac{QoS} $r_k^\mathsf{Min} = r^{\mathsf{Min}} =3$ Mbps, $\forall k \in \mathcal{K}$.

For the allocation of the fronthaul capacities, we distinguish between two regions: 1) The partially-connected regime and 2) the fully-connected regime. In the partially-connected regime, the available fronthaul capacity at each \ac{BS} is insufficient to serve all users from each individual \ac{BS}. For this reason, the allocation of the fronthaul capacities is determined based on the \ac{QoS} $r^{\mathsf{Min}}$ of the users. More precisely, we assume that $r^\mathsf{Min}$ is an integer divisor of $C_n$ in this region. To elaborate, a total fronthaul capacity of $C=21$, for example, results in an allocation of users, so that one \ac{BS} is able to serve one more user that the others, e.g., $[C_{1},C_{2},C_{3}]=[9,6,6]$. The rationale behind this approach is to enable the maximum number of \ac{CoMP} transmissions within the network, given the available resources, because the utilization of this strategy is able increase the \ac{EE} of the network substantially in this regime. To better illustrate the impact of the \ac{CoMP} transmissions, we introduce the \ac{LoSC} as a metric, which captures the number of BS-to-user links within the network. Note that due to the non-zero \ac{QoS} constraints, each user is required to be served by at least one BS.
At the \ac{TP}, $C^{\mathsf{TP}} = N K \, r^\mathsf{Min}$ (here $C^\mathsf{TP}=54$), however, the available resources becomes sufficient to allocate enough resources to each \ac{BS} such that each individual \ac{BS} is able to serve every user with $r^\mathsf{Min}$. It is therefore at the \ac{TP}, where the transition into the fully-connected regime takes place. In the fully-connected regime we assume a symmetric allocation of the fronthaul capacities, namely $C_n=C/N, \forall n\in\mathcal{N}$.

To aid the \acp{BS} in their transmissions, an \ac{IRS} is assumed to be deployed in the center of the area, which consists of $R=15$ reflecting elements. The users and \acp{BS} are positioned uniformly and independently within the operation area. The channel coefficients between the \acp{BS}, users and \ac{IRS} are subject to the standard path-loss model consisting of three components: 1) path-loss as $PL_{x,y} = 148.1 + 37.6\log_{10}(d_{x,y})$, where $d_{x,y}$ is the distance between device $x$ and device $y$ in km; 2) log-normal shadowing with 8 dB standard deviation and 3) Rayleigh channel fading with zero-mean and unit-variance. The channel bandwidth is set to $B=10$ MHz and the noise power spectrum is set to -169 dBm/Hz. Furthermore, the number of Gaussian randomizations is set to $G=25$ and the trade-off parameter is set to $\rho=0.9$, which facilitate a high detection rate for feasible rank-one solutions when solving problem (\ref{eq:genV-alpha_rank}). The maximum number of successive decoding layers at the users are set to $D=K$, as the number of layers and the decoding order at each user is determined dynamically (we choose $D=2$ for the baseline static CMD set selection). The dynamic \ac{CMD} set thresholds are set to $\epsilon^\mathsf{CMD}=-0.4$ and $\epsilon^\mathsf{Thr}=-0.5$, which means that the users, whose \ac{SINR} of potentially decoding another user's message are at least 60\% of the current \ac{SINR} of decoding the message at the intended user, are considered candidates to this user's \ac{CMD} set and are accepted, if their \ac{SINR} is still at 50\% when the \ac{SIC} order at this user is considered. We assume the maximum transmit power per \ac{BS} $P^\mathsf{Tr}$ to be $35$ dBm, the signal processing circuitry of the network is set to $P^\mathsf{circ} = 37$ dBm. Moreover we assume $P^\mathsf{IRS}_r = 10$ dBm and $P^\mathsf{Mbps}=0.3$ W/Mbps. In addition to the dynamic clustering algorithm, we also consider a static clustering algorithm \cite[Algorithm 3]{FH_YU} as baseline clustering approach, which can also be extended accordingly to support the \ac{RS} scenario \cite{AlaaRSCMD}.

\subsection{Impact of the dynamic clustering on the performance}
\begin{figure}
    \begin{center}
   \includegraphics[width=\linewidth]{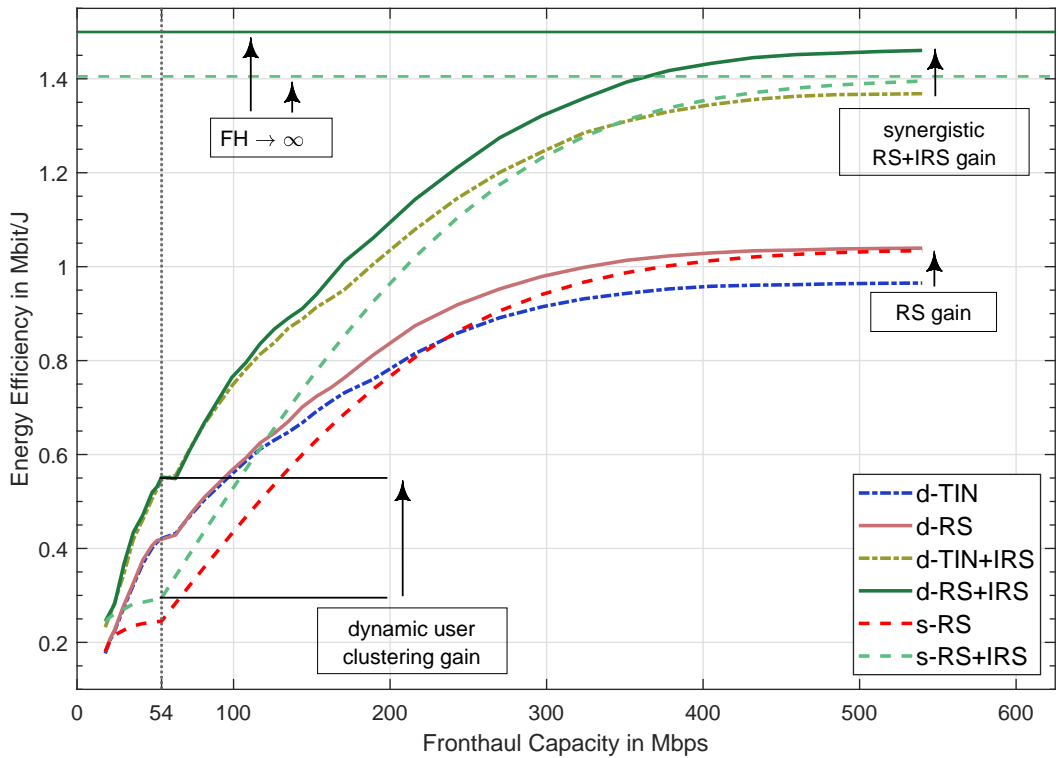}
   \end{center}
   \caption{The system efficiency of all studied schemes, where the prefix ``d-''  represents the dynamic and the prefix ``s-'' the static user clustering scheme. The horizontal lines represent the equivalent broadcast channel scenarios of s-RS+IRS and d-RS+IRS, in which the fronthaul capacity is unlimited. Note, that the d-RS+IRS scheme additionally employs the dynamic \ac{CMD} set allocation. The dynamic \ac{CMD} set allocation obtains sets, which enable synergistic benefits between \ac{RS} and the \ac{IRS}.}
   \label{fig:syseff}
\end{figure}

\begin{figure}
    \begin{center}
   \includegraphics[width=0.9\linewidth]{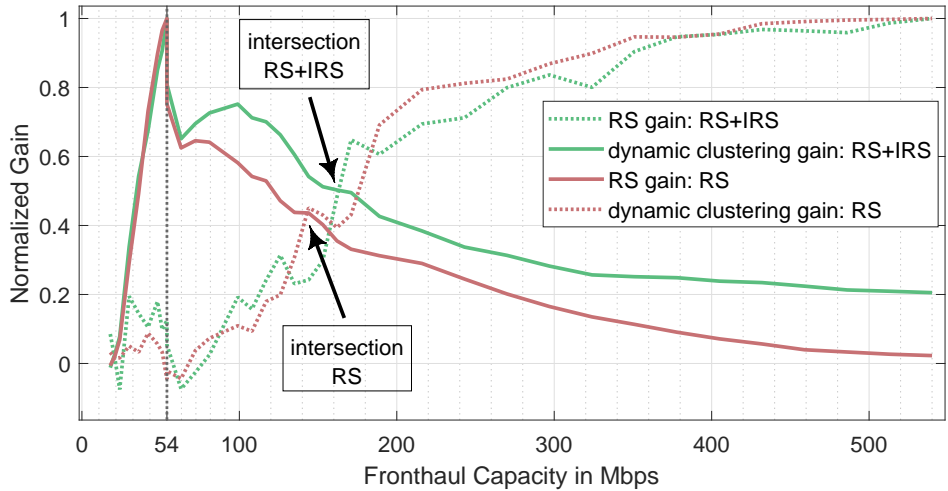}
   \end{center}
   \caption{The normalized gains of \ac{RS} and the dynamic user clustering for the d-RS and d-RS+IRS scenarios. The intersection of the techniques aligns with the local maximum of the peak in Figure \ref{fig:Rc_perc} of the respective scenario.}
   \label{fig:gains}
\end{figure}

First the performance of the dynamic clustering is evaluated against the static clustering in a setup with and without an \ac{IRS}, respectively. Figure \ref{fig:syseff} depicts the system efficiency of the network as a function of the available fronthaul capacity. For the dynamically clustered scenarios, represented by solid lines, we also take the baseline scheme of \ac{TIN} (no common rates) into consideration in order to highlight the gains between the respective techniques. The figure shows that in the low fronthaul capacity region, the performance of the network is mainly limited by the capacity of the fronthaul links. With the scarcity of the fronthaul resources becoming the bottleneck of the system, the efficient clustering of the users to the \acp{BS} becomes the most impactful degree of freedom of the network, especially in the partially-connected regime. This is also depicted in Figure \ref{fig:gains}, showing that the gains of the dynamic clustering are most pronounced at the \ac{TP}. The rationale behind this behaviour is that the dynamic clustering algorithm is able to remove most of the redundancy within the fronthaul transmissions at the TP, resulting in a decrease in the \ac{LoSC} due to the removal of inefficient BS-to-user links. The static clustering, however, is only adding redundancy to the fronthaul transmissions up to the \ac{TP} and is therefore unable to increase the total rate of the network. This results in gains of up to 88\% in energy efficiency for the \ac{IRS}-assisted scenario and up to 71\% for the non-\ac{IRS} case, if compared to the respective statically-clustered scenarios.

\begin{figure}
    \begin{center}
   \includegraphics[width=0.85\linewidth]{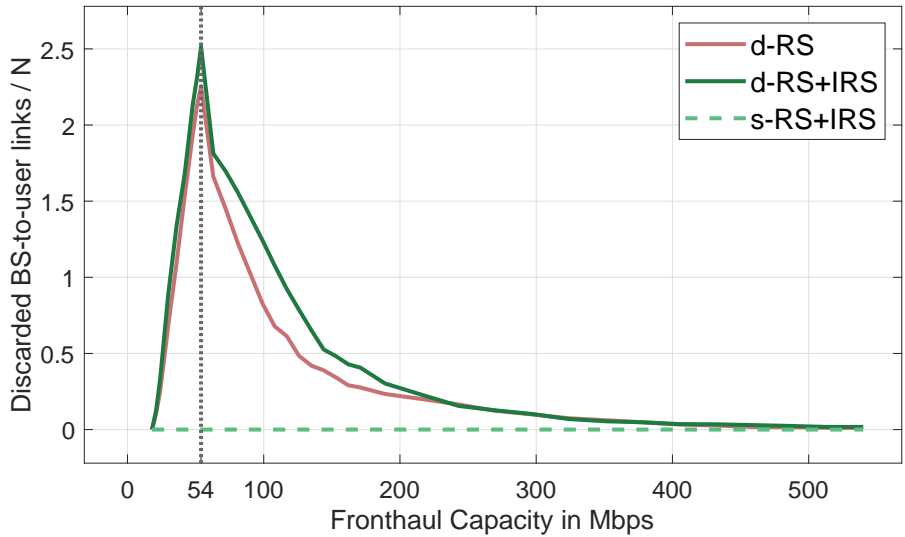}
   \end{center}
   \caption{The decrease in the level of supportive connectivity (\ac{LoSC}) as a function of the fronthaul capacity. The algorithm is able to discard more BS-to-user links in the \ac{IRS}-assisted scenario, which is caused by the increased channel gains of the \ac{IRS}. This leads to a synergistic effect between the dynamic user clustering and the \ac{IRS}.}
   \label{fig:remUsers}
\end{figure}

Note that the gains of the dynamic clustering in the \ac{IRS}-assisted scenario are particularly pronounced as the changes of the phase shifts at the \ac{IRS} during the optimization process can be taken into account, which leads to a synergistic interaction between the \ac{IRS} and the dynamic clustering. More precisely, each reflect element is only able to induce the same phase shift to all the reflected channel paths. With an increasing number of transmissions, the efficiency of the individual phase shifts are decreasing and the complexity of the optimization problem is increasing. Therefore, the feasible set of phase shifters is directly limited by the amount of reflections each reflect element contributes to the network. By removing inefficient BS-to-user links from the network, the number of reflections at the reflect element decreases, causing the feasible set of phase shifters to increase. This enables the selection of more efficient phase shifters, increasing the overall channel gains to some users, which in turn increases the gap between the individual link-efficiencies and encourages the removal of more inefficient links. The effects of this behaviour can also be observed in Figure \ref{fig:remUsers}, which shows that the \ac{IRS}-assisted scenario is able to discard more BS-to-user links than the non-\ac{IRS} scenario on average, particularly in the lower fronthaul capacity region.
\subsection{Impact of the dynamic \ac{CMD} set allocation on the performance}
As we move towards the higher capacity regions, we note that the gains of the dynamic user clustering in Figure \ref{fig:syseff} decrease, while the gains of the \ac{RS}-assisted schemes improve. Especially in the interference-limited regime, the gains of the \ac{RS} technique become more pronounced. This is expected as the \ac{RS} scheme is specifically designed to mitigate interference, hence the increase in its performance. Interestingly, the importance of the dynamic clustering becomes less significant at higher available fronthaul capacities. This causes the curves, which represent the \ac{RS}-enabled non-\ac{IRS} scenarios with the statically-clustered (s-RS) and dynamic-clustered user allocations (d-RS), to converge to the same value. In other words, the performance gain of \ac{RS} over \ac{TIN} at higher fronthaul capacities is solely dependent on the quality of the \ac{CMD} sets. The figure shows two important facts for the \ac{IRS}-assisted scenarios in that regard: 1) the curve, representing the static \ac{CMD} set allocation (s-RS+IRS) does not converge to the same point as the curve, representing the dynamic \ac{CMD} set allocation (d-RS+IRS), but in fact converges at a much lower value; 2) the gain of the (dynamic) \ac{CMD} sets over the \ac{TIN} case is higher in the \ac{IRS}-assisted scenario than in the non-\ac{IRS} assisted scenario.

In essence, by using the dynamic \ac{CMD} sets allocation, the algorithm is able to determine sets, which are of higher quality and more importantly, interacting synergistically with the \ac{IRS} (d-RS+IRS), as opposed to the statically determined \ac{CMD} sets (s-RS+IRS), which only show a marginal improvement over the respective \ac{TIN} scenario (d-TIN+IRS). This synergistic behaviour is highlighted by arrows on the right-hand side of Figure \ref{fig:syseff}, which illustrate the gains of the statically and dynamically determined \ac{CMD} sets. We can see that the arrow representing the gain of the dynamic \ac{CMD} set allocation is larger than the arrow of the statically allocated counterpart, indicating a synergistic interaction between \ac{RS} with the dynamic \ac{CMD} sets and the \ac{IRS}. In fact, if the fronthaul capacity goes to infinity (equivalent to the broadcast channel), the gain between the d-RS+IRS scheme gets even more pronounced if compared to the s-RS+IRS scheme.
\subsection{The Role of Rate Splitting}
\begin{figure}
    \begin{center}
   \includegraphics[width=0.85\linewidth]{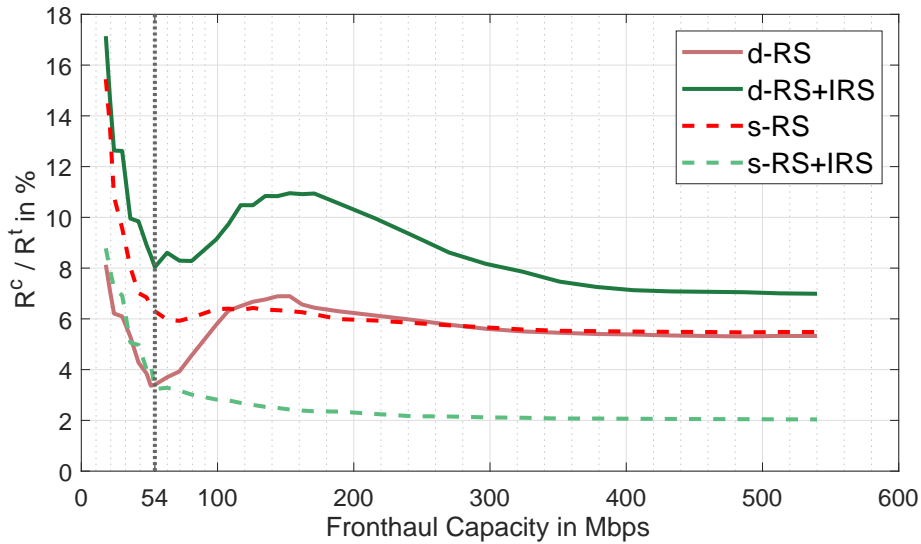}
   \end{center}
   \caption{The proportion of the common rate in the total achievable network rate $R^{\mathsf{t}}$. Both, the d-RS+IRS and the s-RS+IRS case display a peak after the fronthaul capacity overcomes the \ac{TP} (at $C=54$), which is caused by the interaction between the \ac{RS} and dynamic user allocation technique.}
   \label{fig:Rc_perc}
\end{figure}

To illustrate the impact of \ac{RS} and the selected \ac{CMD} sets, we plot the achievable sum rate of the common rate $R^c = \sum_{k\in \mathcal{K}} R^c_k$ in proportion to the total achievable rate $R^t$ of the network in Figure \ref{fig:Rc_perc}. Note, that the impact of the dynamic user clustering decreases in the high fronthaul capacity region, which causes the s-RS and d-RS curves to converge to the same value. However, by comparing the s-RS curve with the d-RS curve in the low fronthaul capacity region, we can see a decrease in the proportion of the common rates within the network. This implies that the dynamic user clustering algorithm mainly detects the BS-to-user allocations of the common message transmissions as inefficient and removes the corresponding BS-to-user links. Moreover, the figure also shows that the proportion of the common rate is decreasing, as the fronthaul capacity increases towards the \ac{TP} for all schemes. This signifies, that using \ac{CoMP} transmissions to reduce the required transmit power within a fronthaul-limited network is more efficient than the utilization of \ac{RS} in this regime, which fits with the fact that \ac{RS} performs best in the interference-limited regime. In opposition to the {d-RS} scheme, the {d-RS+IRS} scheme counteracts the removal of inefficient common-rate links by finding sensible \ac{CMD} sets dynamically. The more efficient sets result in an overall \textit{higher} common rate percentage, especially in the low fronthaul regime, if compared to the s-RS+IRS case and also causes the convergence to a higher value, representing the higher quality of the determined sets. Interestingly, the curves, which represent the dynamic user allocation schemes, are characterized by a peak that begins right after the \ac{TP}. This peak is a result of the interaction between the two techniques that are employed in these scenarios, namely \ac{RS} and the dynamic user clustering.

In fact, as the fronthaul increases after the \ac{TP}, the dynamic clustering gain decreases, while the effectiveness of the rate splitting increases. This leads to an intersection, around which both techniques are able to interact with each other. Therefore, it is also around this intersection, where the user allocations, representing the common rate links, get more efficient than the respective private allocations, while the clustering algorithm is still able to remove these inefficient private links. To illustrate this behaviour, we plot the normalized gains of the individual techniques as a function of the fronthaul capacity in Figure \ref{fig:gains}. The figure shows that the intersection, at which the gains of the individual techniques meet, overlaps with the maximum of the respective peak in Figure \ref{fig:Rc_perc}. It is also at this point, where the growth of the energy efficiency of the system (see Figure \ref{fig:syseff}) for the d-RS+IRS scenario increases significantly.

%% file: sections/conclusion.tex
\section{Conclusion}\label{ch:conc}
Modern communication networks are forecast to require an increased amount of transmit power, as more users and \ac{BS} connect to the network, causing an increase in interference. Motivated by these predictions, this work studies the deployment of a passive \ac{IRS} in a \ac{C-RAN} with the aim to increase the energy efficiency.
To mitigate the additional increase in interference, caused by the \ac{IRS}, the impact of the interaction between the \ac{RS-CMD} scheme and the \ac{IRS} on the \ac{EE}, under the assumption of finite fronthaul capacities, is studied. 
An optimization problem is formulated, which includes a dynamic user clustering and a dynamic \ac{CMD} set allocation, in order to take the changes of the phase shifts throughout the optimization process into account. As the optimization variables are highly dependent on each other, the problem is split into three subproblems, which are solved in an alternating fashion. A strategy is devised, in which the outer loop is able to influence the alternating optimization by regulating outer-loop parameters, resulting in high quality solutions to the optimization problem. 

Numerical results show that the deployment of the \ac{IRS} in combination with the dynamically-determined \ac{CMD} sets displays a synergistic benefit with regards to the \ac{EE}, if used together with the \ac{RS} technique. In addition to that, the dynamic clustering allocation is also displaying a synergistic interaction with the \ac{IRS} resulting in an increased gain of 88\% over the gain of 71\% for the non-\ac{IRS} scenario. The results, moreover, show that with an increasing available fronthaul capacity, the gain of the dynamic clustering decreases, while the gain of \ac{RS} simultaneously increases. Around the resulting intersection, the growth of the energy efficiency of the network increases notably as both, the dynamic user clustering and the dynamic \ac{CMD} set allocation technique, are able to interact with each other. The proposed methods offer an attractive upgrade path to existing networks as these can be easily and cost efficiently extended by deploying an \ac{IRS} in a RS-enabled \ac{C-RAN} to realize future communication networks.

%% file: content/acronyms.tex
\begin{acronym}
\setlength{\itemsep}{0.1em}
\acro{AF}{amplify-and-forward}
\acro{AWGN}{additive white Gaussian noise}
\acro{B5G}{Beyond 5G}
\acro{BS}{base station}
\acro{C-RAN}{Cloud Radio Access Network}
\acro{CMD}{common-message-decoding}
\acro{CM}{common-message}
\acro{CoMP}{coordinated multi-point}
\acro{CP}{central processor}
\acro{D2D}{device-to-device}
\acro{DC}{difference-of-convex}
\acro{EE}{energy efficiency}
\acro{IC}{interference channel}
\acro{i.i.d.}{independent and identically distributed}
\acro{IRS}{intelligent reflecting surface}
\acro{IoT}{Internet of Things}
\acro{LoS}{line-of-sight}
\acro{LoSC}{level of supportive connectivity}
\acro{M2M}{Machine to Machine}
\acro{MIMO}{multiple-input and multiple-output}
\acro{MRT}{maximum ratio transmission}
\acro{MRC}{maximum ratio combining}
\acro{NLoS}{non-line-of-sight}
\acro{PSD}{positive semidefinite}
\acro{QCQP}{quadratically constrained quadratic programming}
\acro{QoS}{quality-of-service}
\acro{RF}{radio frequency}
\acro{RS-CMD}{rate splitting and common message decoding}
\acro{RS}{rate splitting}
\acro{SDP}{semidefinite programming}
\acro{SDR}{semidefinite relaxation}
\acro{SIC}{successive interference cancellation}
\acro{SCA}{successive convex approximation}
\acro{SINR}{signal-to-interference-plus-noise ratio}
\acro{SOCP}{second-order cone program}
\acro{SVD}{singular value decomposition }
\acro{TP}{transition point}
\acro{TIN}{treating interference as noise}
\acro{UHDV}{Ultra High Definition Video}
\acro{LoSC}{level of supportive connectivity}

\end{acronym}